\documentclass[10pt,onecolumn]{autart}    % Enable this line and disable the 
\usepackage{graphicx}          % Include this line if your 
\usepackage[dvips]{epsfig}    % or this line, depending on which
\usepackage{amsmath,amssymb,amsfonts} % assumes amsmath package installed
\usepackage{mathrsfs}
\usepackage{pstricks}
\usepackage{verbatim}
\usepackage{url}
\usepackage{subfigure}
\usepackage{latexsym}
\usepackage{epsf}
\usepackage{epsfig}
\usepackage{psfrag}
\usepackage{pdfsync,srcltx}
\usepackage{ifthen}
\usepackage{algorithm}
\usepackage{algpseudocode}
\usepackage{epstopdf}
\usepackage{float}
\usepackage{fancyhdr}
\usepackage{kbordermatrix}

\DeclareFontFamily{OT1}{pzc}{}
\DeclareFontShape{OT1}{pzc}{m}{it}{<-> s * [1.200] pzcmi7t}{}
\DeclareMathAlphabet{\mathpzc}{OT1}{pzc}{m}{it}
\newcommand{\G}{\mathcal{G}}
\newcommand{\graphset}{\mathbb{G}}
\newcommand{\weightset}{\mathbb{W}}
\newcommand{\V}{{\mathpzc{V}}}
\newcommand{\E}{{\mathpzc{E}}}

\newcommand{\R}{\mathbb{R}}

\newcommand{\Z}{\mathbf{Z}}
\newcommand{\PaperORReport}{Report}  

\newcommand{\scr}{\mathcal}
\newcommand{\eqdef}{:=}
\renewcommand{\qed}{\hfill $\square$}

\newcommand{\Prob}{P}
\newcommand{\MarkovP}{\mathcal{P}}
\newcommand{\diag}{\text{diag}}

\DeclareMathOperator{\Exp}{E}
\newcommand{\boldepsilon}{\mbox{\boldmath$\epsilon$}}

\newcommand{\boldmu}{\mbox{\boldmath$\mu$}}

\newcommand{\mbf}{\mathbf}                      % Bold face

\begin{document}

\begin{frontmatter}
%\runtitle{Insert a suggested running title}  % Running title for regular 
                                              % papers but only if the title  
                                              % is over 5 words. Running title 
                                              % is not shown in output.

\title{Estimation from Relative Measurements in Mobile Networks with Markovian Switching Topology: Clock Skew and Offset Estimation for Time Synchronization\thanksref{footnoteinfo}} % Title, preferably not more 
                                                % than 10 words.

\thanks[footnoteinfo]{This work has been supported by the National
    Science Foundation by Grants CNS-0931885 and ECCS-0955023. Author
    email addresses: {\tt\small \{cdliao,pbarooah\}@ufl.edu}.}

%%% first author
\author{Chenda Liao},
\author{Prabir Barooah}
\address{Department of Mechanical and Aerospace Engineering, University of Florida, Gainesville, FL 32611, USA}

%\author[Paestum]{Marcus Tullius Cicero}\ead{cicero@senate.ir},    % Add the 
%\author[Rome]{Julius Caesar}\ead{julius@caesar.ir},               % e-mail address 
%\author[Baiae]{Publius Maro Vergilius}\ead{vergilius@culture.ir}  % (ead) as shown
%
%\address[Paestum]{Buckingham Palace, Paestum}  % Please supply                                              
%\address[Rome]{Senate House, Rome}             % full addresses
%\address[Baiae]{The White House, Baiae}        % here.

                             % keyword list or with the 
                                          % help of the Automatica 
                                          % keyword wizard

\begin{abstract}                          % Abstract of not more than 200 words.
 We analyze a distributed algorithm for estimation of
  scalar parameters belonging to nodes in a mobile network from noisy
  relative measurements. The motivation comes from the problem of
  clock skew and offset estimation for the purpose of time
  synchronization. The time variation of the network was modeled as a
  Markov chain. The estimates are shown to be mean square convergent
  under fairly weak assumptions on the Markov chain, as long as the
  union of the graphs is connected. Expressions for the asymptotic
  mean and correlation are also provided. The Markovian switching
   topology model of mobile networks is justified for certain node
   mobility models through empirically estimated conditional entropy
   measures.
\end{abstract}

\end{frontmatter}

\section{Introduction}\label{sec:intro}
We consider the problem of estimation of variables in a network of
mobile nodes in which pairs of communicating nodes can obtain noisy
measurement of the difference between the variables associated with
them. Specifically, suppose the $u$-th node of a network has an
associated \emph{node variable} $x_u \in \R$. If nodes $u$ and $v$ are
neighbors at discrete time index $k$, then they can obtain a
measurement $\zeta_{u,v}(k)$ where
\begin{align}\label{eq:rela-meas}
  \zeta_{u,v}(k) = x_u - x_v + \epsilon_{u,v}(k).
\end{align}
The problem is for each node to estimate its node variable from the
relative measurements it collects over time, without requiring any
centralized information processing or coordination. We assume that at
least one node knows its variable. Otherwise the problem is
indeterminate up to a constant. A node that knows its node variable is
called a \emph{reference node}. All nodes are allowed to be mobile, so
that their neighbors may change with time.

The problem of time synchronization (also called clock-synchronization) through clock skew and offset
estimation falls into this category, and provides the main motivation
for the study. Time synchronization in ad-hoc networks, especially in wireless sensor
networks, has been a topic of intense study in recent years. The
utility of data collected and transmitted by sensor nodes depend
directly on the accuracy of the time-stamps. In TDMA based
communication schemes, accurate time synchronization is required for
the sensors to communicate with other sensors. Operation on a
pre-scheduled sleep-wake cycle for energy conservation and lifetime
maximization also requires accurate knowledge of a common global
time. We refer the interested reader to the review
papers~\cite{FS_BY_IN:04,Sundararaman_UG_AK_AdHocNetworks:05,BMS_AS_MILCOM:06} for more
details on time synchronization. 

The relationship between local clock time $\tau_u(t)$ of node $u$ and global
time $t$ is usually modeled as
$\tau_u(t) = \alpha_u t +  \beta_u
$,
where the scalars $\alpha_u, \beta_u$ are called its \emph{skew} and
\emph{offset}, respectively~\cite{FS_BY_IN:04,BMS_AS_MILCOM:06}. A node can determine the global time $t$
from its local clock time by using the relationship 
$\hat{t} = (\tau_u(t) -\hat{\beta}_u)/\hat{\alpha}_u$
as long as it can obtain estimates $\hat{\alpha}_u,\hat{\beta}_u$ of
the skew and offset of its local clock. Hence the problem is clock
synchronization in a network can be alternatively posed as the problem
of nodes estimating their skews and offsets. It is not possible for a
node to measure its skew and offset directly. However, it is possible
for a pair of neighboring nodes to measure the difference between their
offsets and logarithm of skews by exchanging a number of time stamped
messages. Existing protocols to perform so-called \emph{pairwise
  synchronization}, such
as~\cite{KN_QC_ES_BS_TC:07,SY_CV_MS_TSN:07,ML_YW_TVT:10},
can be used to obtain such relative measurements. The details will be
described in Section~\ref{sec:measurements}. The problem of clock
offset and skew estimation can therefore be cast as a special case of
the estimation from relative measurements described above. If an
algorithm is available to solve the scalar node variable estimation
problem, nodes can execute two copies of this algorithm in parallel to
estimate both skew and offset. Therefore we only consider the scalar
case. In the context of time synchronization, the existence of a
reference node means that at least one node has access to the global
time $t$. This is the case when at least one node is equipped with a
GPS receiver, in which case that node has access to the UTC
(Coordinated Universal Time). If no node has a GPS receiver, then one node has to be
elected to be the reference so that it's local clock time is considered
the global time that everyone has to synchronize to.

\subsection{Related work}
Time synchronization in sensor networks can be classified into pairwise synchronization and global
synchronization methods. In  pairwise synchronization, a pair of nodes
try to synchronize their clocks to each other. In practice this is
often achieved by one of the nodes estimating its relative offset
and/or skew with respect to the other node, so that the local time of
the  other node serves as a reference~\cite{KN_QC_ES_BS_TC:07,SY_CV_MS_TSN:07,ML_YW_TVT:10,YW_QC_ES_SPM:11}. Precise definitions of
relative offset and relative skew are postponed till
Section~\ref{sec:measurements}. In global synchronization, also called
network-wide synchronization, all nodes synchronize
themselves to a common time. % Fundamental limits of pairwise and global time
% synchronization are discussed in~\cite{Freris_IEEETAC:10}.

A common approach for global synchronization in sensor networks is to
first elect a root node and construct a spanning tree of the network
with the root node being the ``level 0'' node. Every node thereafter
synchronizes itself to a node of lower level (higher up in the
hierarchy) by using a pairwise synchronization method. Examples of such
spanning-tree based protocols include Timing-Sync
Protocol for Sensor Networks (TPSN)~\cite{SG_RK_MS_SenSys:03} and
Flooding Time Synchronization Protocol
(FTSP)~\cite{MM_BK_GS_AL_SenSys:04}. Change in the network topology
due to node mobility or node failure requires recomputing the spanning
tree and sometimes even re-election of the root node. This adds
considerable communication overhead. The situation gets worse if nodes
move rapidly.

Recently, a number of fully distributed global synchronization
algorithms have been proposed that do not need spanning tree
computation. Distributed protocols are therefore more readily
applicable to mobile networks than tree-based protocols. Among the
distributed synchronization protocols proposed, some are based on
estimation of the skew and/or offset of each clock with respect to a
reference clock (called \emph{absolute time
  synchronization}). The algorithms proposed
in~\cite{PB_JH_2ICISIP:05,PB_NdS_JH_LNCS:06,Kumar-timesync-II:06,FN_BV_KP:09,ML:YCW:11}
belong to this category. Another class of protocols estimate a common
global time that may not be related to the time of any clock in the
network. The algorithms proposed
in~\cite{RC_SZ:10,RC_ED_SZ:11,LS_FF_Automatica:11} belong to this
category, which we call \emph{virtual time synchronization}.

\subsection{Contribution}
In this paper we consider the problem of distributed estimation of
skews and offsets with respect to a reference clock in a mobile
network for global absolute time synchronization, where the network
changes with time due to nodes' motion. The common thread among
virtual time synchronization methods mentioned earlier is the use of
consensus-type algorithms to construct virtual skew and offsets that
every node agrees to. In many applications, absolute time
synchronization is preferable over virtual time synchronization. This
occurs when the user of the sensor network is interested in the time
of an event that is measured in an absolute reference time, such as
UTC provided by a GPS unit on a base station.  Therefore, in this
paper we consider only absolute time synchronization.

We analyze an algorithm for estimating absolute skews and offsets from
noisy pairwise relative measurements of skews and offsets, which is a
slight modification of the algorithms proposed
in~\cite{Karp_03TimeSync,PB_JH_2ICISIP:05,Kumar-timesync-II:06}. Though
the algorithm is adopted from these earlier papers, the analysis in
those papers were limited to static networks. Thus, little is known
about how such an algorithm will perform in a mobile network.

The main contribution is that we analyze the
convergence of the algorithm when the network topology changes due to
the motion of the nodes, as well as random communication failure. We
model the resulting time-varying topology of the network as the state
of a Markov chain. Techniques for the analysis of jump linear systems
from~\cite{CostaFragosoMarques:04} are used to study convergence of
the algorithm. We show that under fairly weak assumptions on the
Markov chain, the proposed algorithm is mean square convergent if and
only if the union of the graphs that occur is connected. Mean square
convergence means the expected value and the variance of the estimates
obtained by each node converges to fixed values that do not depend on
the initial conditions. When the relative measurements are unbiased,
then limiting mean is the same as the true value of the variable,
meaning the estimates obtained are asymptotically unbiased. Formulas
for the limiting mean and variance are obtained by utilizing results
from jump linear systems.

The algorithm we analyze bears a close resemblance to consensus
algorithms. In fact, the estimation error dynamics turns out to be a
leader-follower consensus algorithm, where the leader states -
corresponding to the estimation error of the reference nodes - are
always $0$. However, existing results from consensus cannot be
directly used to analyze the scenario examined in this
paper. Consensus literature almost always treats the problem where all
nodes participates in the consensus algorithm, i.e., ``leaderless
consensus'', while ours is a ``leader-follower'' consensus since the
reference nodes error state stays at $0$. One may expect analysis of
this case would be easier, but that turns out to be not the case. Even
though the literature on consensus is extensive,
%  (see
% ~\cite{KS_MJMF:09,MH_DS_NG_HJ:10,TL_JFZ:10,JL_XL_WX:11,LC_YZ_QZ:11,YFS_JH:12}
% and the references therein)
the topic of consensus with both time-varying graph topology and
additive measurement noise is considered only in a limited number of
papers,
e.g.~\cite{KS_MJMF:09,MH_DS_NG_HJ:10,TL_JFZ:10,JL_XL_WX:11}. There are
several differences between the consensus algorithms studied
in~\cite{KS_MJMF:09,MH_DS_NG_HJ:10,TL_JFZ:10,JL_XL_WX:11} and the
error dynamics examined in this paper, which preclude using their
results to perform the analysis. These include requirement of symmetry
or balance in graphs/matrices, preassignment of time-varying gains
that must be synchronized among all nodes, etc. None of these
restrictions are imposed in our analysis (see
Remark~\ref{rem:consensus} for more details). 
%In addition, use of
%Markov jump linear systems theory allows us to provide formulas for
%limiting mean and correlation. In~\cite{CL_PB_automatica_arXiv:13}, we
%show that when nodes move according to the Random Waypoint Mobility
%model~\cite{TC_JB_VD_WCMC:02}, the resulting switching dynamics can
%indeed be modeled by a Markov chain. This provides justification for
%assuming the switching of topology is Markovian when analyzing mobile
%networks.

%Although the papers~\cite{RC_SZ:10,RC_ED_SZ:11,LS_FF_Automatica:11}
%propose distributed protocols for time synchronization, those
%protocols are not based on estimation of absolute skew and offset of
%the nodes, which is the problem we examine. Thus, their algorithms are
%outside the scope of our analysis.

Another contribution of the paper is to provide justification for the
Markovian switching topology for mobile networks. The Markovian
switching model has also been used extensively in studying consensus
protocols in networks with dynamic
topologies~\cite{VG_BH_RM_CDC:03,YZ_YT_Automatica:09,SK_JM_TSP:10,MH_DS_NG_HJ:10}. For
a network of static nodes with link drops, the Markovian switching
model arises naturally from Markovian link drop model. In mobile
networks, though, the only case where we can prove that a mobile
network evolves according to a Markov chain is when nodes move
according to the so-called random walk mobility
model~\cite{TC_JB_VD_WCMC:02}. Although the Markovian switching
assumption facilitates analysis, this assumption requires
justification for more complex motion models. We use a technique
from~\cite{CC:73} to check if the graph switching is Markovian if
nodes according to the so-called Random Waypoint Mobility (RWP)
model. The RWP model is one of the most widely used mobility models
for ad-hoc mobile networks~\cite{TC_JB_VD_WCMC:02}. We show that the
resulting graph switching process can indeed be approximated well by a
(first order) Markovian switching model.

A preliminary version of this paper was presented
in~\cite{CL_PB_CDC:10}. Compared to that paper, we make several
additional contributions. While the paper~\cite{CL_PB_CDC:10} provided
only sufficient conditions for mean square convergence, here we
provide both necessary and sufficient conditions. An assumption of
symmetry of certain matrices were made in~\cite{CL_PB_CDC:10}, which
is removed in the present paper. 

The rest of the paper is organized as
follows. Section~\ref{sec:prob-form} describes the
connection between the problem of estimation from relative
measurements and the problem of skew/offset estimation, and then states the problem precisely. Section~\ref{sec:main-section}
 describes the proposed algorithm and
states the main result (Theorem~\ref{thm:ms-convergence-timevarying}). It also discusses the relevance of
the Markovian switching topology model. Section~\ref{sec:proof} is devoted to the
proof of the theorem. Simulation studies are presented in
Section~\ref{sec:simulations}.

\section{The estimation problem}\label{sec:prob-form}
We consider the problem of estimating the scalar parameters (called \emph{node
variables}) $x_u$, $u=1,\dots,n_b$, where $n_b$ is the number of nodes in
the network that do not know their node variables. We assume that
there are $n_r$ additional nodes that knows their node variables,
where $n_r \geq 1$. These define a node set
$\V=\{1,\dots,n\}$, where $n = n_b+ n_r$ is the total number of nodes. For later reference, we define $\V_b
\eqdef \{1,\dots,n_b\}$ and $\V_r=\{n_b+1,\dots,n_b + n_r\}$, so that
$\V = \V_b \cup \V_r$. Note that $n=n_b+n_r$. Time is measured by a
discrete time-index $k=0,1,\dots$. The mobile nodes define a
time-varying undirected \emph{measurement graph} $\G(k) = (\V,\E(k))$,
where $(u,v) \in \E(k)$ if and only if $u$ and $v$ can obtain a
relative measurement of the form~\eqref{eq:rela-meas} during the time
interval between the time indices $k$ and $k+1$. Specifically, for
each $(u,v)\in \E(k)$, there is a measurement $\zeta_{u,v}(k) = x_u -
x_v + \epsilon_{u,v}(k)$ that is available to both $u$ and $v$ at time
$k$. In practice, one of the two nodes computes this measurement from
sensed information. We assume that if $u$ computes the measurement
$\zeta_{u,v}$, it then sends this measurement to $v$ so that $v$ also
has access to the same measurement. We follow the convention that the
relative measurement between $u$ and $v$ that is obtained by the node
$u$ is always of $x_u-x_v$ while that used by $v$ is always of $x_v -
x_u$. Since the same measurement is shared by a pair of neighboring
nodes, if $v$ receives the measurement $\zeta_{v,u}$ from $u$, then it
converts the measurement to $\zeta_{v,u}$ by assigning $\zeta_{v,u}(k)
\eqdef -\zeta_{u,v}(k)$. We assume, without any loss of generality, that between
a pair of nodes $u$ and $v$, the node with the lower index obtains the
relative measurement between them first, and then shares with the node
with the higher index. % This technical assumption is made to
% ensure that the statistics of the realative measurement does not
% change with time between the same pair of nodes. 

The \emph{neighbors} of $u$ at $k$, denoted by $\scr{N}_u(k)$, is the
set of nodes that $u$ has an edge with in the measurement graph
$\G(k)$. We assume that if $v \in \scr{N}_u(k)$, then $u$ and $v$ can
also exchange information through wireless communication at time
$k$. Therefore, if one prefers to think of a communication graph, we
assume that it is the same as the measurement graph.

The task is to estimate the node variables $x_u$ for $u=1,\dots,n$
by using the relative measurements $\zeta_{u,v}(k), (u,v)\in \E(k)$
that becomes available over time $k=0,1,\dots$. In addition, the
algorithm has to be distributed in the sense that each node has to
estimate its own variables, and at every time $k$, a node $u$ can only
exchange information with its neighbors $\scr{N}_u(k)$. Note that the
estimation problem is indeterminate unless $n_r>0$.

\subsection{Relation to skew and offset estimation}\label{sec:measurements}
% As explained in Section~\ref{sec:intro}, global time synchronization in a
% network can be cast equivalently as the problem of each node $u$
% estimating its skew and offset ($\alpha_u,\beta_u$) with respect to a
% reference clock. First we explain how it is possible to use existing
% methods of pairwise synchornization to obtain relative measurements of
% log-skews and offsets, before formaly defining the estimation problem. 
 
To see the connection between skew/offset estimation and the problem
of estimation from noisy relative measurements introduced in the
previous section, we first discuss the notion of pairwise
synchronization between a pair of neighboring nodes $u$ and $v$. By
exchanging a number of time-stamped messages, it is possible for node
$u$ to estimate the so-called \emph{relative skew} $\alpha_{u,v}$ and
\emph{relative offset} $\beta_{u,v}$ between itself and
$v$, where
\begin{align}\label{eq:relative-skew-offset}
  \tau_{u}(t) = \alpha_{u,v}\tau_v(t) + \beta_{u,v}.
\end{align}
That is, the parameters $\alpha_{u,v}$ and $\beta_{u,v}$ relate the
local time of $u$ to the local time of $v$ at the same global time
$t$. A number of methods are available that allows
pairwise synchronization between a node pair from time-stamped
messages~\cite{KN_QC_ES_BS_TC:07,SY_CV_MS_TSN:07,LN_ES_KQ_TWC:08,FN_BV_KP:09,ML_YW_TVT:10}. The
parameters $\alpha_{u,v}$ and $\beta_{u,v}$ are also referred to as
the \emph{skew and offset of node $u$ with respect to node
  $v$}~\cite{YW_QC_ES_SPM:11}.

The relationship between the absolute skew and offset 
$\alpha_u,\beta_u,\alpha_v,\beta_v$ and relative skew and offset
$\alpha_{u,v},\beta_{u,v}$ is given by
\begin{align}\label{eq:skew-ratio}
\alpha_{u,v} & \eqdef \frac{\alpha_u}{\alpha_v} &&& \beta_{u,v} & \eqdef
\beta_u - \beta_v \frac{\alpha_u}{\alpha_v}.
\end{align}
This relationship is obtained by expressing the local time $\tau_u(t)$ of node
$u$ at global time $t$ in terms of the local time $\tau_v(t)$ at node $v$ at the
same time $t$ by using~\eqref{eq:rela-meas}:
\begin{align*}
\tau_{u}(t) & = \alpha_u(\frac{\tau_v (t) - \beta_v}{\alpha_v}) + \beta_u=
\frac{\alpha_u}{\alpha_v}\tau_v(t) + \beta_u - \beta_v \frac{\alpha_u}{\alpha_v},  
\end{align*}
and comparing with~\eqref{eq:relative-skew-offset}. Suppose a node $u$ obtains noisy estimates
$\hat{\alpha}_{u,v},\hat{\beta}_{u,v}$ of the parameters
$\alpha_{u,v},\beta_{u,v}$ by using a pairwise synchronization
protocol. 
\begin{enumerate}
\item We model the noisy estimate of $\alpha_{u,v}$ as
\begin{align}\label{eq:alphaij-hat}
  \hat{\alpha}_{u,v} & = \exp(\epsilon_{u,v}^s)\alpha_{u,v}
\end{align}
where $\exp(\cdot)$ is exponential function and $\epsilon_{u,v}^s$ is a random variable. If the
estimation error is small, then $\epsilon_{u,v}^s$ is close to
$0$. Taking log, we get  
\begin{align}\label{eq:log-alpha-hat}
\log \hat{\alpha}_{u,v} & = \log \alpha_{u,v} + \epsilon_{u,v}^s = \log \alpha_u -  \log \alpha_v + \epsilon_{u,v}^s .
\end{align}
Eq.~\eqref{eq:log-alpha-hat} can be rewritten as 
$\zeta_{u,v}^s = x_u^s - x_v^s + \epsilon_{u,v}^s$,
with the definitions
$\zeta_{u,v}^s\eqdef \log \hat{\alpha}_{u,v}$ and $x_{i}^s \eqdef \log \alpha_u$,
which makes $\zeta_{u,v}^s$ a noisy relative measurement of the node
variables $x_u^s$ and $x_v^s$; cf.~\eqref{eq:rela-meas}. It is
important to notice that $\zeta_{u,v}^s$ is a measured quantity --
since $\hat{\alpha}_{u,v}$ is measured -- while the variables
$x_u^s,x_v^s$, which are logarithms of the skews, are unknown.
\item Similarly, the noisy
estimate $\hat{\beta}_{u,v}$ of $\beta_{u,v}$ with random estimation error $e_{u,v}^o$ can be written as
\begin{align}\label{eq:beta-hat}
  \hat{\beta}_{u,v} & = \beta_{u,v} + e_{u,v}^o = \beta_u - \beta_v + \epsilon_{u,v}^o,
\end{align}
where
$\epsilon_{u,v}^o \eqdef \beta_v(1 - \alpha_u/\alpha_v) + e^o_{u,v}$.
Again,~\eqref{eq:beta-hat} can be rewritten as 
$\zeta_{u,v}^o = x_u^o - x_v^o + \epsilon_{u,v}^o$,
with the definitions
$\zeta_{u,v}^o\eqdef \hat{\beta}_{u,v}$ and $x_{u}^o \eqdef \beta_u$,
which makes $\zeta_{u,v}^o$ a noisy relative measurements of the node
variables $x_u^o$ and $x_v^o$; cf.~\eqref{eq:rela-meas}. In this case the
node variables are the clock offsets $\beta_u$'s. The noise $\epsilon_{u,v}^o$ in the offset
measurement is in general biased even if the measurement of the relative
offset $\beta_{u,v}$ is unbiased. 
\end{enumerate}
This discussion shows that the estimates of the relative skew and the
relative offset between a pair of neighboring nodes, which can be
obtained by existing algorithms for pairwise synchronization, can be
expressed as a noisy relative measurement of node variables by
appropriate redefinitions. The node variables are log-skews and
offsets. Once node $u$ obtains estimates 
$\hat{x}_u^s$ and $\hat{x}_u^o$
of its two node variables $x_u^s$ and $x_u^o$, it can estimate its
skew and offset as $\hat{\alpha}_u \eqdef \exp(\hat{x}_u^s)$ and $\hat{\beta}_u \eqdef \hat{x}_u^o$. Thus, the problem of estimating the skews and
offsets of all the clocks in a network can be transformed to an estimation
from relative measurements problem, where relative measurements are of
the form~\eqref{eq:rela-meas}.

\begin{rem}\label{rem:parallel}
  From this point on, we only consider the estimation problem
  involving scalar node variables. This entails no loss of generality
  since estimation of the two scalar variables, skew and offset, can
  be performed in parallel. In the skew estimation problem, log-skews
  take the role of node variables and $\log \hat{\alpha}_{u,v}$'s
  obtained from pairwise synchronization take the role of relative
  measurements. In the offset estimation problem, node variables are
  the offsets and relative measurements are the  $ \hat{\beta}_{u,v}$'s obtained from pairwise
  synchronization. The assumption on the existence of the
  reference node is equivalent to at least one node knowing the
  global time. This can be achieved by either one or more nodes
  having access to GPS time, or by arbitrarily
  electing a node as a reference and choosing its local time as the
  global time.
\end{rem}

% If the skews are equal to unity, then it is possible to obtain a noisy
% measurement of the difference between their offsets with an additive
% zero mean noise from time stamped measurements.  In fact, the
% papers~\cite{Kumar-timesync-II:06} make such assumptions and use that
% to justify modeling the noise in both forms of relative measurements
% as zero mean. The question of bias of the estimates are ignored. One
% reason we go throgh the detailed derivation of the relative
% measurements is to expose that offset measurements will have bias, so
% the zero mean assumption of the previous papers is not always
% appropriate. This in turn justifies examining the bias of the
% estimates.

\section{Algorithm and results}\label{sec:main-section}
\subsection{Algorithm for distributed estimation from relative measurement}
The algorithm we consider is adopted
from~\cite{PB_JH_2ICISIP:05,Kumar-timesync-II:06,Karp_03TimeSync}, with minor
modification to make it applicable to time varying networks.  Each
node $u$ maintains in its local memory an estimate $\hat{x}_u(k)$ of
its node variable $x_u \in \R$. Every node - except the reference
nodes - iteratively updates its estimate as we'll describe now. The
estimates can be initialized to arbitrary values. In executing the
algorithm at iteration $k$, node $u$ communicates with its current
neighbors to obtain measurements $\zeta_{u,v}(k)$ and their current
estimates $\hat{x}_v(k)$, $v \in \scr{N}_u(k)$. Since obtaining
measurements require exchanging time-stamped messages, the current
estimates can be easily exchanged during the process of obtaining new
measurements. Node $u$ then updates its estimate according to
\begin{align}\label{eq:algo}
\hat{x}_u(k+1)= 
\left\{ \begin{array}{rcl}
\frac{w_{uu}(k)\hat{x}_u(k) + \sum_{v\in
      \scr{N}_u(k)}w_{vu}(k)(\hat{x}_v(k) +
    \zeta_{u,v}(k))}{w_{uu}(k)+\sum_{v\in
      \scr{N}_u(k)}w_{vu}(k)} , &  u\in\V_b \\ 
x_u, & u\in\V_r,
\end{array}\right.
\end{align}
where the \emph{weights} $w_{vu}(k)$ and $w_{uu}(k)$ are arbitrary positive numbers.  The update law is well-defined even at times when $u$ has no neighbors. Nodes continue this iterative update unless they see little change in their local
estimates, at which point they can stop updating. The update procedure
in each node $u\in\V_b$ is specified in Algorithm~\ref{alg:sync_imp}.

\begin{algorithm}                      % enter the algorithm environment
\caption{Distributed update at node $u$}% give the algorithm a caption
\label{alg:sync_imp}                           % and a label for \ref{} commands later in the document
\begin{algorithmic}[1]                    % enter the algorithmic environment
\State{Initialize estimate $\hat{x}_u(0) \in \R$ and local iteration counter $k=0$}
\While{$u$ is performing iteration}
        \If{$\scr{N}_u(k_u)\neq \emptyset$}
        \For{$v \in \scr{N}_u(k)$}
            \If{$u$ does not have $\zeta_{u,v}(k)$} 
                 \State{1.$u$ and $v$ perform pairwise
                   synchronization: $u$ obtains $\zeta_{u,v}(k)$, $v$ obtains $\zeta_{v,u}(k)$};
                \State{2.$u$ and $v$ exchange their current
                  estimates: $u$ saves $\hat{x}_v(k)$; $v$ saves $\hat{x}_u(k)$};
             \Else
             \State{$u$ does not communicate with $v$};
             \EndIf 	
        \EndFor 	  
                 \State{$u$ updates $\hat{x}_u(k+1)$ using
                   \eqref{eq:algo}};
          \Else
	     \State{$\hat{x}_u(k+1) \leftarrow\hat{x}_u(k)$};
	     \EndIf       
        \State {$k$=$k+1$};                
\EndWhile
\end{algorithmic}
\end{algorithm}

Each node $u$ is allowed to vary its local weights $w_{uv}(k)$ with
time and use distinct weights for distinct neighbors to account for
the heterogeneity in measurement quality. Between two neighbors $p,q$
of node $u$ at time $k$, the relative measurement $\zeta_{u,p}(k)$ between
$u$ and $p$ may have lower measurement error than the relative
measurement $\zeta_{u,q}(k)$ between $u$ and $q$. This occurs, for
example, if $u$ and $p$ were able to exchange more time stamped
messages than $u$ and $q$ before computing the relative
measurements~\cite{ML_YW_TVT:10,YW_QC_ES_SPM:11}. In this case, node
$u$ should choose its local weights at $k$ so that
$w_{pu}(k)>w_{qu}(k)$. Due to the denominator in~\eqref{eq:algo}, it
is only the ratios among the weights that matter, not their absolute
values.

\subsubsection{Asynchronous implementation}\label{sec:async-implement}
The description so far is in terms of a common global iteration index
$k$. In practice, nodes do not have access to such a global
index. Instead, each node keeps a local \emph{iteration index}. After every increment of the local index, the node tries to
collect a new set of relative measurements with respect to one or more
of its neighbors within a pre-specified time interval. At the end of
the time interval, whether it is able to get new measurements or not,
it updates its estimate according to the update law~\eqref{eq:algo} and increments its local iteration
counter. Now the index $k$ in~\eqref{eq:algo} has to be interpreted as
the local iteration index. The process then repeats. It follows from~\eqref{eq:algo}
that if a node is unable to gather new measurements from any
neighbors, then its updated estimate is precisely the previous estimate.

The global iteration index is useful to describe the algorithm from
the point of view of an omniscient spectator. Let $T$ the time
interval, say, in seconds, between two successive increments of the
global index $k$. The parameter $T$ is arbitrary, as long as is small
enough so that no node updates its local estimate more than once with
the time interval $T$.  In that case, one of only two events are
possible for an arbitrary node $u$ at the end of the time interval
when the global counter is increased from $k$ to $k+1$: (i) $u$ either
increases its local index by one, or (ii) $u$ does not increases its
local index. If a node increases its local index, both the local and
global indices increase by one.  A node does not increase its local
iteration index if it is not able to gather new measurements. In the
omniscient spectator's view, the node's neighbor set is empty at this
time index; so according to~\eqref{eq:algo}, the next estimate of the
node's variable is the same as the previous one. Thus, a node's local
asynchronous state update can be described in terms of the synchronous
algorithm~\eqref{eq:algo}; the latter being more convenient for
exposition. We therefore consider only the synchronous version in the
sequel.

\subsection{Convergence analysis with Markovian switching}
In this paper we model the sequence of measurement
graphs $\{\G(k)\}_{k=0}^{\infty}$ that appear as time progresses as
the realization of a (first order) Markov chain, whose state space
$\graphset =\{\G_1,\dots,\G_N\}$ is the set of graphs that can occur
over time.  % In general, a stochastic process $X(k)$ is called
% $\ell$-th order Markov if $\Prob(X(k+1) | X(k),X(k-1),\dots) =
% \Prob(X(k+1) | X(k),X(k-1),\dots,X(k-\ell+1))$, where $\Prob(\cdot)$
% denotes probability. 
The Markovian switching assumption on the graphs
means that $\Prob(\G(k+1) = \G_i | \G(k)=\G_j) = \Prob(\G(k+1) = \G_i
| \G(k)=\G_j,\G(k-1)=\G_\ell,\dots,\G(0)=\G_p)$ where $\G_i,
\G_j,\G_\ell,\dots,\G_p\in \graphset$ and where $\Prob(\cdot)$
 denotes probability. We assume that the Markov chain is homogeneous,
and denote the transition probability matrix of the chain by
$\MarkovP$, in which $p_{ij}$ is the $(i,j)$-th entry of $\MarkovP$. Further
discussion on Markov modeling of graphs is postponed till Appendix~\ref{sec:markov_modeling}. 

%\begin{rem}\label{rem:Markov}
%  The examination of the applicability of the Markovian model of graph
%  switching is provided in~\cite[ Appendix
%  A]{CL_PB_automatica_arXiv:13}. We examine the Random Waypoint
%  Mobility model by using an empirical conditional entropy based
%  method suggested in~\cite{CC:73}. The analysis
%  in~\cite{CL_PB_automatica_arXiv:13} reveals that the graph switching
%  process can be modeled as a (first order) Markov chain.
%\end{rem}
Let $e_u(k) \eqdef \hat{x}_u(k)-x_u$ be the estimation error at node
$u$. Since $\zeta_{u,v}(k) = x_u - x_v + \epsilon_{u,v}(k)$, the
update law~\eqref{eq:algo} can be rewritten as
\begin{align}\label{eq:algo-error}
e_u(k+1) = 
\left\{ \begin{array}{rcl}
\frac{w_{uu}(k)e_u(k) + \sum_{v\in
      \scr{N}_u(k)}w_{vu}(k)(e_v(k) +
    \epsilon_{u,v}(k))}{w_{uu}(k)+\sum_{v\in
      \scr{N}_u(k)}w_{vu}(k)} & \mbox{for}
& u\in\V_b \\ 
0 & \mbox{for} & u\in\V_r,
\end{array}\right.
\end{align}
The right hand side
of~\eqref{eq:algo-error} is a weighted average of estimation errors of
$x_u$ and measurement noise. If the measurement noise
$\epsilon_{u,v}(k)$ is zero-mean and the initial estimates are
unbiased, i.e. $\Exp[e_u(0)] = 0$, $\forall u\in\V_b$, then
$\Exp[e_u(k)] = 0$ for all $k$, where
$\Exp[\cdot]$ denotes expectation.

The main result of the paper - on the mean square convergence of~\eqref{eq:algo-error} - is stated below as a theorem. In the statement of
theorem, $\mbf e(k) \eqdef [e_1(k),\dots,e_{n_b}(k)]^T$ is the estimation
error vector. Moreover, $\boldmu(k)\eqdef\Exp[\mbf
e(k)]$ is the mean and $\mathbf{Q}(k)\eqdef \Exp[\mbf
e(k){\mbf e(k)}^T]$ is the correlation matrix of the
estimation error vector. We say that a stochastic process $\mbf{y}(k)$ is \emph{mean square convergent} if $\Exp[\mbf{y}(k)]$ and 
$\Exp[\mbf{y}(k) \mbf{y}^T(k)]$ converges as $k\to\infty$ for every initial condition. The
\emph{union graph} $\hat{\G}$ is defined as follows:
\begin{align}\label{eq:G-union}
  \hat{\G} \eqdef \cup_{i=1}^{N}\G_i = (\V,\cup_{i=1}^{N}\E_i),
\end{align}
where $\E_i$ is set of edges in $\G_i$. We assume that the measurement
noise $\epsilon_{u,v}(k)$ affecting the measurements on the edge
$(u,v)$ is a wide sense stationary process. We also assume that the measurement noise sequence
$\epsilon_{u,v}(k)$ and the initial condition $\hat{x}_u(0)$, for any
$u,v,k$ is independent of the Markov chain that governs the
time-variation of the graph.

Due to technical reasons, we make an additional assumption that there
exists a time $k_0$ after which the edge-weights do not change. % That is, if
% $u$ and $v$ become neighbors at two distinct times $k>k_0$, then
% $w_{uv}(k)$ and $w_{vu}(k)$ keep constant. Similarly, for each node
% $u$, $w_{uu}(k)$ keeps constant for $k>k_0$.
The choice of weights during the transient period (up to $k_0$) will
affect initial reduction of the estimation errors but will not change
the asymptotic behavior.

Recall that $\mbf e(k)$ is the estimation
error vector for the nodes who do not know their node variables, the main theorem is as follows:

\begin{thm}\label{thm:ms-convergence-timevarying}
Assume that the temporal evolution of the communication graph $\G(k)$ is governed by
an $N$-state homogeneous Markov chain that is ergodic, and $p_{ii}>0$ for
$i=1,\dots,N$. The estimation error $\mbf e(k)$ is mean square convergent if and only if $\hat{\G}$ is
connected.  
\end{thm}

\begin{rem}\label{rem:unbiased}
  The formulas for computing the limiting values $\lim_{k \to \infty}
  \boldmu(k)$ and $\lim_{k \to \infty} \mathbf{Q}(k)$ are provided in
  Lemma~\ref{lem:ss-var} (Section~\ref{sec:proof}). It follows
  directly from the formulas (see Lemma~\ref{lem:ss-var}), that if additionally
  all the measurements are unbiased, then $\lim_{k \to \infty}
  \boldmu(k)= 0$.
\end{rem}

The implication of the theorem is that as long as nodes are connected
in a ``time-average'' sense characterized by $\hat{\G}$ being
connected, the estimates of the node variables will converge to random
variables with a constant mean and variance, irrespective of the initial conditions. Thus, after
a sufficiently long time, the nodes can turn off the synchronization
updates without much loss of accuracy. The assumption of ergodicity of
the Markov chain ensures that there is an unique steady state
distribution and that the steady state probability of each state is
non-zero~\cite{CostaFragosoMarques:04}. This means every graph in the
state space of the chain occurs infinitely often. Since their union
graph is connected, ergodicity implies that information from
the reference node(s) will flow to each of the nodes over time. None of the
graphs that ever occur is required to be a connected graph. The
assumption $p_{ii}>0$ means $P(\G(k+1)=\G_i|\G(k)=\G_i)>0$. This can
be assured if the nodes move slowly enough.

\begin{rem}\label{rem:consensus}[Relation to consensus]
  The estimation error dynamics~\eqref{eq:algo-error} can be
  interpreted as a ``leader-following'' consensus algorithm, where the
  state of node $u$ is the estimation error $e_u(k)$ for $u \in \V_b$,
  while the leader states are $e_r(k) \equiv 0$ for $r\in\V_r$.
  Although the literature on consensus is extensive, the topic of
  consensus with time-varying graph topology and additive measurement
  noise is considered only in a limited number of papers,
  with~\cite{KS_MJMF:09,MH_DS_NG_HJ:10,TL_JFZ:10,JL_XL_WX:11}
  representing the state of the art in this topic. There are
  significant differences between the algorithm we analyze and those
  in~\cite{MH_DS_NG_HJ:10,TL_JFZ:10}, as well as between the
  results. First, the cited references deal with the leaderless
  consensus while our situation is that of a leader-following
  one. Second. the algorithms in~\cite{MH_DS_NG_HJ:10,TL_JFZ:10}
  require that the nodes use a specifically designed time-varying
  weight sequence that satisfy a certain \emph{persistence} condition:
  : they decay to $0$ while being square summable but not absolutely
  summable. That is $\{w_{u,v}(k)\}_{1}^{\infty} \in \ell_2, \notin
  \ell_1$ for each pair $(u,v)$. This condition is difficult to ensure
  unless the nodes have synchronized clocks to begin with. In
  contrast, we allow the nodes to vary their weights with time
  arbitrarily subject only to the condition that they stop doing so at
  a certain time.  Furthermore, the results
  in~\cite{KS_MJMF:09,MH_DS_NG_HJ:10,TL_JFZ:10} are established under
  the assumption the weighted Laplacian matrices\footnote{The
    Laplacian matrix of $\G_i$ is equal to $M_i-N_i$ in this paper,
    where the definition of matrices $M_i$ and $N_i$ are given in
    Section~\ref{sec:proof}.}  of the directed graphs are
  balanced. Ensuring balanced weights require coordination between
  pairs of neighbors. In contrast, we do not impose any kind of
  symmetry on the Laplacian matrices, so that each node can choose its
  weights without coordinating with its neighbors. Not imposing
  symmetry makes the analysis significantly more difficult.
\end{rem}

\section{Proof of Theorem~\ref{thm:ms-convergence-timevarying}}\label{sec:proof}
\begin{figure}
\begin{center}
\psfrag{Gw1}{$\G$}
\psfrag{Gw2}{$\vec{\G}$}
\psfrag{w1}{$w_{12}$}
\psfrag{w2}{$w_{21}$}
\psfrag{w3}{$w_{32}$}
\psfrag{w4}{$w_{23}$}
\psfrag{w5}{$w_{31}$}
\psfrag{w6}{$w_{13}$}
\psfrag{n1}{$w_{11}$}
\psfrag{n2}{$w_{22}$}
\psfrag{n3}{$w_{33}$}
\includegraphics[scale = 0.5, clip = true]{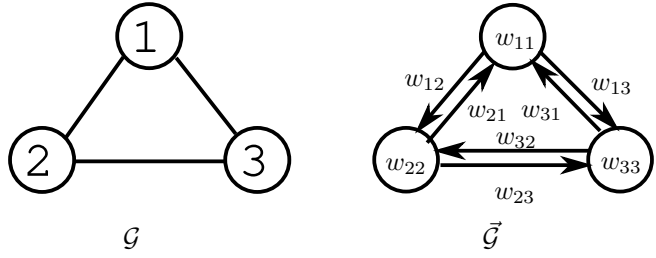}
\caption{A measurement graph $\G$ and the corresponding weight graph
  $\vec{\G}$.}
\label{fig:com_dg_comp}
\end{center}
\end{figure}

% We first introduce some standard and non-standard graph-theoretic
% terminology. 
We consider a \emph{weighted directed} graph $\vec{\G}(k) =
(\V,\vec{\E}(k),W(k))$ associated with undirected measurement graph $\G(k) = (\V,\E(k))$. In particular, there exists an undirected edge (u,v) in $\G(k)$, then there exist
two directed edges $(u,v)$ and $(v,u)$ in $\vec{\G}(k)$. The \emph{weight matrix} $W(k)$ defined as 
\begin{align}
W_{uv}(k)\eqdef \left\{ \begin{array}{rcl}
w_{vu}(k)>0 & \mbox{for}
& (u,v)\in\vec{\E}(k) \\ 
w_{uu}(k)>0 & \mbox{for} & v=u \\
0 &  & \text{o.w.}
\end{array}\right.
\end{align}
Thus, given a
measurement graph $\G(k)$ and $W(k)$, $\vec{\G}(k)$ is specified. See Figure~\ref{fig:com_dg_comp} for an example of
an undirected measurement graph and an associated directed weighted
graph. 
The square \emph{non-negative} matrices, $D(k)$, $M(k)$ and $N(k)$ is defined as follows: $D(k)$ is a $n\times n$ diagonal matrix made up of the diagonal entries of $W(k)$, and $N(k)\eqdef W(k)-D(k)$. $M(k)$ is a $n\times n$ diagonal matrix with entry $M_{uu}(k)=\sum_{u\neq v}W_{uv}(k)$. Furthermore, we define the $n_b \times n_b$ \emph{basis matrix} $D_b(k)$, $M_b(k)$ and $N_b(k)$ as the principle submatrix of $D(k)$, $M(k)$ and $N(k)$ obtained by removing those rows and columns
corresponding to the reference nodes% , i.e., from $(n_b+1)$-th to $n$-th rows and columns
. Now,~\eqref{eq:algo-error} can be compactly expressed as 
\begin{align}\label{eq:JLS-input}
\mbf{e}(k+1) & = J_b(k)\mbf e(k) + B_b(k)\boldepsilon(k),
\end{align}
where
\begin{equation}\label{eq:J-theta-defn}
\begin{split}
&J_b(k) \eqdef (M_b(k)+D_b(k))^{-1}(N_b(k)+D_b(k)),\\
&B_b(k) \eqdef  (M_b(k)+D_b(k))^{-1}A_b(k),\\
&\boldepsilon(k)\eqdef [\bar{\epsilon}_1(k)^T,\dots,\bar{\epsilon}_{n_b}(k)^T]^T,\\ &\bar{\epsilon}_u(k)\eqdef[\epsilon_{u,1}(k),\dots  \epsilon_{u,n}(k)]^T, \\
& A_b(k)\eqdef diag(\bar{N}_1(k),\dots,\bar{N}_{n_b}(k)),\\&  \bar{N}_u(k)\eqdef [N_{u1}(k),\dots ,N_{un}(k)],
\end{split}
\end{equation}
where vector $\bar{\epsilon}_u(k)$ and $\bar{N}_u(k)$ do not contain
$\epsilon_{u,u}(k)$ and $N_{uu}(k)$ respectively. Note that diagonal
matrix $M_b(k)+D_b(k)$ is always non-singular because diagonal entries
in $D_b(k)$ are always positive as $W_{uu}(k)>0$ for all $u$, and
$M_b(k)$ is nonnegative. When $N_{uv}(k)=0$, the corresponding
$\epsilon_{u,v}(k)$ is taken to be an arbitrary random variable with
mean and variance such that the stationary assumption is
satisfied. Since these noise terms are multiplied by $0$, this entails
no loss of generality. Moreover, recall that
$\epsilon_{u,v}(k)=-\epsilon_{v,u}(k)$.

% In general, the weight $W(k)$ at time $k$ is not completely specified
% by the graph $\G(k)$ at that time if nodes are allowed to vary the weights arbitrarily over
% time. However, recall that an additional constraint was imposed on
% choosing weights, that there exists a time $k_0$ after which the
% weight between two nodes do not change. 
As a result of the assumption that there exists a time $k_0$ after
which the weight between two nodes do not change, the graph $\G(k)$
uniquely determines the weight matrix $W(k)$ for $k>k_0$. Since there
are $N$ distinct graphs in $\graphset$, a set $\weightset \eqdef
\{W_1,\dots,W_N\}$ is also defined, with $W_i$ associated with $\G_i$.
As a result, for $k \geq k_0$, if $\G(k) = \G_i$ then
$W(k)=W_i$. Therefore, $D_i,M_i,N_i,D_{bi},M_{bi},N_{bi},J_{bi}$ are
uniquely defined by $\{\G_i,W_i\}$.

With these choices stated above, the state of the
following system is identical to that of~\eqref{eq:JLS-input} for the
same initial conditions:
\begin{align}\label{eq:JLS-input-gen}
  \mbf e(k+1) &= J_{b\; \theta(k)} \mbf e(k)+B_{b\; \theta(k)}
  \boldepsilon(k), \quad k \geq k_0
\end{align}
where $\theta: \Z^+ \to \{1,\dots,N\}$ is the switching process that is
governed by the underlying Markov chain $\G(k)$. The reason for the qualifier
$k \geq k_0$ is that weights are not limited to the set $\weightset$
before $k_0$, so technically the matrices $J_{b\; \theta(k)}$ and $B_{b\; \theta(k)}$ are uniquely
determined by the Markov chain only for $k\geq k_0$.
The error dynamics~\eqref{eq:JLS-input-gen} is a Markov jump linear system
(MJLS)~\cite{CostaFragosoMarques:04}.  To proceed with the analysis of the mean square convergence of \eqref{eq:JLS-input-gen}, we need some terminology.
\begin{align}\label{eq:1}
  \gamma & \eqdef \Exp [\boldepsilon(k)], & \Gamma & \eqdef
  \Exp[\boldepsilon(k)\boldepsilon^T(k)], \\
\boldmu(k) & \eqdef \Exp[\mbf{e}(k)], & \mbf{Q}(k) & \eqdef \Exp[\mbf{e}(k)\mbf{e}^T(k)].
\end{align}
Furthermore, for a set of matrices
$X_i \in \R^{\ell_1 \times \ell_2}$, $Y_{ij} \in \R^{\ell_1 \times
  \ell_2}$, $i,j=1,\dots,N$, denote the $\ell_1 N \times \ell_2 N$ block diagonal matrix $\diag[X_i]=\diag\{X_1,\dots.X_N\}$ and 
\begin{align*}
[Y_{ij}] \eqdef
\begin{bmatrix}
Y_{11} &  \dots & Y_{1N} \\
\vdots & \ddots & \vdots \\
Y_{N1} &  \dots & Y_{NN}
\end{bmatrix}_{\ell_1 N \times \ell_2 N}. 
\end{align*}
Now, define the matrices 
\begin{align}\label{eq:def_J_bar}
  \begin{split}
J_i  & \eqdef (M_i+D_i)^{-1}(N_i+D_i) \in \R^{n \times n}, \\    
 J_{bi}  &\eqdef (M_{bi}+D_{bi})^{-1}(N_{bi}+D_{bi}) \in \R^{n_b\times n_b},\\
F_i & \eqdef
J_i\otimes J_i \in \R^{n^2 \times n^2}, \; F_{bi}  \eqdef J_{bi}\otimes J_{bi} \in \R^{n_b^2\times n_b^2}
\end{split}
\end{align}
where $\otimes$ denotes the Kronecker product. Furthermore, define the matrices
\begin{align}
\mathcal{D} & \eqdef \left( \MarkovP^{T} \otimes I
\right)diag[F_i]  =  [p_{ji}F_j]  \in \R^{N n^2 \times N
  n^2},  \label{eq:define_D}\\
 \mathcal{D}_b &\eqdef(\MarkovP^T\otimes I)diag[F_{bi}]  =
 [p_{ji}F_{bj}]  \in
 \mathbb{R}^{Nn_b^2 \times Nn_b^2}, \label{eq:D-bar-def}\\
\mathcal{C}_b &\eqdef(\MarkovP^T\otimes I)diag[J_{bi}]  =
[p_{ji}J_{bj}]    \in \mathbb{R}^{Nn_b
  \times Nn_b}, \nonumber
 \end{align}
where $I$ is an identity matrix of appropriate dimension. Recall that
$\MarkovP$ is the transition probability matrix of the Markov chain.

\medskip

The key to establish Theorem~\ref{thm:ms-convergence-timevarying}, is the  following
technical result and the proof is provied in the Appendix~\ref{proof-lemma:rho_of_D} since it
requires introduction of considerable new terminology.

\begin{lem}\label{lemma:rho_of_D}
When the temporal evolution of the graph $\G(k)$ is governed by a
homogeneous ergodic 
Markov chain whose transition probability matrix $\MarkovP$ has the property that
its diagonal entries are strictly positive, then $\rho(\mathcal{D}_b)<1$ if and only if the union graph
$\hat{\G}$ defined in~\eqref{eq:G-union} is connected, where
$\mathcal{D}_b$ is defined in~\eqref{eq:D-bar-def} and $\rho(\cdot)$
denotes the spectral radius. If $\hat{\G}$ is not connected, $\rho(\mathcal{D}_b)=1$.
\end{lem}

%%%%%%%%%%%%%%%%%%%%% formulas for mean and variance from JLS %%%%%%%%%%
The following definitions and terminology
from~\cite{CostaFragosoMarques:04} will be needed in the sequel.  Let $\mathbb{R}^{\ell_1 \times \ell_2}$ be
the space of $\ell_1\times \ell_2$ real matrices. Let $\mathbb{H}^{\ell_1 \times \ell_2}$ be the set of all
N-sequences of real $\ell_1\times \ell_2$ matrices, so that $V \in \mathbb{H}^{\ell_1
  \times \ell_2}$ means $V =(V_1,V_2,\dots,V_N)$ where $V_i\in \mathbb{R}^{\ell_1\times
  \ell_2}$ for $i=1,\dots,N$.  The operators $\varphi$ and $\hat{\varphi}$ is defined to
create a tall vector by stacking together columns from these matrices,
as follows: let ${(V_i)}_j\in \R^{\ell_1}$ be the $j$-th column of $V_i \in
\R^{\ell_1 \times \ell_2}$, then
\begin{align}\label{eq:tall_matrix}
 \varphi(V_i) & \eqdef [(V_i)_1^T,\dots,(V_i)_n^T]^T\in \mathbb{R}^{\ell_1\ell_2}\\
\hat{\varphi}(V) &\eqdef[\varphi(V_1)^T,\dots,\varphi(V_N)^T]^T\in\mathbb{R}^{N\ell_1\ell_2}.
\end{align}
Similarly, the inverse function $\hat{\varphi}^{-1}: \R^{N\ell_1\ell_2} \to
\mathbb{H}^{\ell_1 \times \ell_2}$ is defined so that it produces an element of
$\mathbb{H}^{\ell_1 \times \ell_2}$ given a vector in $\R^{N\ell_1\ell_2}$. 
% \begin{lemma}\label{lem:ss-var}
%   Consider the jump linear system~\eqref{eq:JLS-input-gen} with an
%   underlying homogeneous and ergodic Markov chain. The state vector
%   $\mbf{e}(k)$ of the system~\eqref{eq:JLS-input-gen} converges in the
%   mean square sense if and only if $\rho(\mathcal{D}_b)<1$, where
%   $\mathcal{D}_b$ is defined in~\eqref{eq:D-bar-def}. When mean square
%   convergence  occurs, then $\boldmu(k) \to \boldmu$ and $\mbf{Q}(k) \to \mbf{Q}$, where
% \begin{align}\label{eq:mu-Q}
%   \boldmu & \eqdef\displaystyle\sum_{i=1}^{N} q_i ,\quad
% \mathbf{Q} \eqdef\displaystyle\sum_{i=1}^{N} Q_i,
% \end{align}
% where
% \begin{align*}
% [q_1^T,\dots,q_N^T]^T &= q \eqdef (I-\mathcal{C}_b)^{-1}\psi
% \quad \in \R^{Nn_b},\\
% (Q_1,\ldots,Q_N) &= Q \eqdef \hat{\varphi}^{-1}\left(
%   (I-\mathcal{D}_b)^{-1}\hat{\varphi}(R(q))\right) \quad  \in \mathbb{H}^{n_b \times n_b} ,
% \end{align*}
% and $\psi,R(q)$ are given by
% \begin{align*}
% \psi &\eqdef [\psi_1^T, \dots, \psi_N^T]^T
% \in \mathbb{R}^{Nn_b} , ~~ \psi_j  \eqdef \displaystyle\sum_{i=1}^{N}p_{ij}B_{bi}\gamma\pi_i \in \R^{n_b},\\
%   R(q) &\eqdef  (R_1(q),\ldots,R_N(q))\in \mathbb{H}^{n_b \times n_b},\\
%  R_j(q)&\eqdef  \displaystyle\sum_{i=1}^{N}p_{ij}(B_{bi}\Gamma B^T_{bi}\pi_i+J_iq_i
% \gamma^T B^{T}_{bi} +B_{bi}\gamma q_i^TJ_i^T))\in \mathbb{R}^{n_b \times n_b}.
% \end{align*}
% Moreover, $\mbf{Q}$ is positive semi-definite.
% \end{lemma}
\begin{lem}\label{lem:ss-var}
Consider the jump linear system~\eqref{eq:JLS-input-gen} with an underlying homogeneous and ergodic Markov chain. The state vector $\mbf{e}(k)$ of the system~\eqref{eq:JLS-input-gen} converges in the mean square sense if and only if $\rho(\mathcal{D}_b)<1$, where $\mathcal{D}_b$ is defined in~\eqref{eq:D-bar-def}. When mean square convergence occurs, then $\boldmu(k) \to \boldmu$ and $\mbf{Q}(k) \to \mbf{Q}$, where
\begin{align}\label{eq:mu-Q}
\boldmu & \eqdef\displaystyle\sum_{i=1}^{N} q_i ,\quad 
\mathbf{Q} \eqdef\displaystyle\sum_{i=1}^{N} Q_i,
\end{align}
where
\begin{align*}
[q_1^T,\dots,q_N^T]^T &= q \eqdef (I-\mathcal{C}_b)^{-1}\psi \quad \in \R^{Nn_b},\\
(Q_1,\ldots,Q_N) &= Q \eqdef \hat{\varphi}^{-1}\left((I-\mathcal{D}_b)^{-1}\hat{\varphi}(R(q))\right) \in \mathbb{H}^{n_b \times n_b},
\end{align*}
and $\psi,R(q)$ are given by
\begin{align*}
&\psi \eqdef [\psi_1^T, \dots, \psi_N^T]^T \in \mathbb{R}^{Nn_b} , ~ \psi_j \eqdef \displaystyle\sum_{i=1}^{N}p_{ij}B_{bi}\gamma\pi_i \in \R^{n_b},\\
&R(q)\eqdef (R_1(q),\ldots,R_N(q))\in \mathbb{H}^{n_b \times n_b},\\
&R_j(q)\eqdef \displaystyle\sum_{i=1}^{N}p_{ij}(B_{bi}\Gamma
B^T_{bi}\pi_i+J_{bi}q_i \gamma^T B^{T}_{bi} +B_{bi}\gamma q_i^TJ_{bi}^T))\in \mathbb{R}^{n_b \times n_b}.
\end{align*}
Moreover, $\mbf{Q}$ is positive semi-definite.
\end{lem}
\begin{pf*}
 {Proof:} It follows from Theorem $3.33$, Theorem $3.9$, and  remark
 3.5 of~\cite{CostaFragosoMarques:04} that mean square convergence
  of~\eqref{eq:JLS-input-gen} is equivalent to
  $\rho(\mathcal{D}_b)<1$. 
% \footnote{For the interested reader, the matrix
%     $\mathcal{D}_b$ is referred to as $\mathcal{A}_1$
%     in~\cite{CostaFragosoMarques:04}.}.  
  The expressions for the mean and correlation, as well as the fact
  that $\mbf{Q}\geq 0$, also follow from~\cite[Proposition
  3.37,3.38]{CostaFragosoMarques:04}.  The existence of the steady
  state distribution $\pi$ (that appear in the formulas) follows from
  the ergodicity of the Markov chain. \qed
\end{pf*}
%%%%%%%%%%%%%%%

%%%%%%%%%%%%%%%%% proof of the theorem %%%%%%%%%%%%%%%%%%%
Now we are ready to prove
Theorem~\ref{thm:ms-convergence-timevarying}.

\begin{pf*}{Proof of Theorem~\ref{thm:ms-convergence-timevarying}}  
  \textbf{(Sufficiency)}: It follows from the hypotheses and Lemma~\ref{lemma:rho_of_D}
  that we have $\rho(\mathcal{D}_b)<1$. It then follows from
  Lemma~\ref{lem:ss-var} that the state converges in the mean square
  sense. % The statement about the asymptotic mean being unbiased if the
%   measurement noise is zero mean follows immediately from the
%   expression for the liming mean in Lemma~\ref{lem:ss-var} by plugging
%   in $\gamma=0$.
  \textbf{(Necessity)}: If the union of graph is not
  connected, we have from Lemma~\ref{lemma:rho_of_D} that
  $\rho(\mathcal{D}_b)=1$. This shows that (due to
  Lemma~\ref{lem:ss-var}) convergence will not occur.\qed
\end{pf*}

%
%\begin{remark}\label{rem:consensus}
%  The equation \eqref{eq:JLS-input} can be interpreted as a
%  leader-following consensus problem, where the consensus state for
%  node $u$ is the estimation error $e_u(k)$ for $u \in \V_b$, while the leader states are $e_r(k) \equiv 0$ for $r\in\V_r$. Consensus problem are
%  more interesting when there is no leader. Results from leaderless
%  consensus can be used to achieve convergence of consensus with
%  leaders, as remarked in~\cite{MH_DS_NG_HJ:10}. The
%  papers~\cite{MH_DS_NG_HJ:10,TL_JFZ:10} propose leaderless
%  consensus algorithm using stochastic approximation under time-varying graphs, in which node variables converge to a common value in mean square sense. However, they require all weighted Laplacian matrices of directed graphs, i.e.  $M_i-N_i$ in this paper to be balanced or satisfied another very restricted condition. In addition, measurement noise is modeled as Martingale difference sequence (therefore zero mean). Consequently,  the
%  algorithms proposed in~\cite{MH_DS_NG_HJ:10,TL_JFZ:10} cannot be applied to achieve mean square convergence
%under assumptions in this paper. \newstuff{\cite{LM_TAC:05} proposes a leaderless consensus (node variables to a common value) method under time-varying graph as well. The weighted average algorithm proposed there is similar to that in this paper, but measurement of neighbors' states are under noiseless assumption.}   
%\end{remark}

\section{Simulation studies}\label{sec:simulations}
As discussed in Section~\ref{sec:measurements}, skew and offset
estimation are special cases of the problem of estimation of scalar
node variables from relative measurements. Therefore simulations are conducted
only for scalar node variable estimation. In all simulations, node
variables are chosen arbitrarily, a single reference node is present,
and the value of the its node variable is $0$. The noise on each
measurement is a normally distributed random variable. All the edge
weights are assigned a value of unity at every time.

\subsection{Four-node network with Markovian switching}\label{sec:simulations_4nodes}
\begin{figure}
\begin{center}
\includegraphics[scale = 0.5]{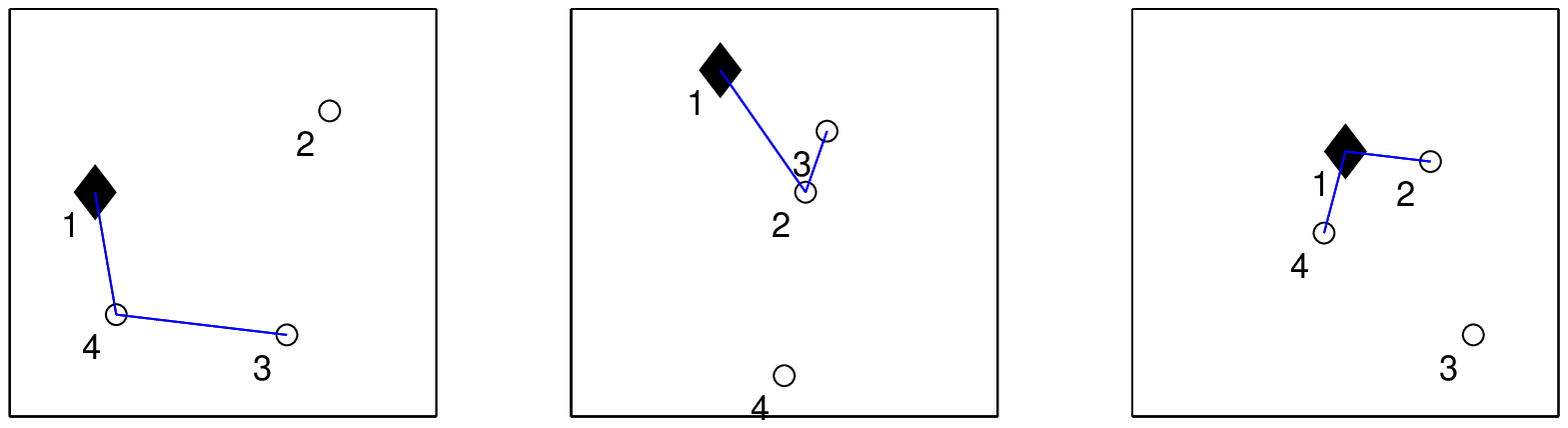}
\caption{The three graphs $\G_1,\G_2,\G_3$ that comprises $\graphset$. Node $1$ is the
  reference.}
\label{fig:hist_graph_4nodes_spe_P}
\end{center}
\end{figure}

In this scenario the nodes move in such a way that the graph
$\G(k)$ can be one of only $3$ graphs shown in
Figure~\ref{fig:hist_graph_4nodes_spe_P}. The graphs change according to
a Markov chain whose transition probability matrix is
 \begin{align}\label{eq:P-test}
    \MarkovP = \begin{bmatrix}
0.3 & 0 & 0.7\\
                    0.1 & 0.5 & 0.4\\
                    0    & 0.5 & 0.5
\end{bmatrix}. 
 \end{align}
 Notice that none of the graphs is a connected graph, though the union
 of the graphs in $\graphset$ is connected. Also, $\MarkovP$ is
 ergodic. The mean and variance of measurement noise on every edge are
 chosen as $0$ and $10^{-4}$, respectively. The limiting means and
 variances of the estimates therefore can be computed from the
 predictions of Lemma~\ref{lem:ss-var}. Monte-Carlo experiments are
 conducted to empirically estimate the mean and variance of the
 estimation error, by averaging over $1000$ sample runs.
\begin{figure}
\centering
 \subfigure[Mean]{
  \includegraphics[scale = 0.5]{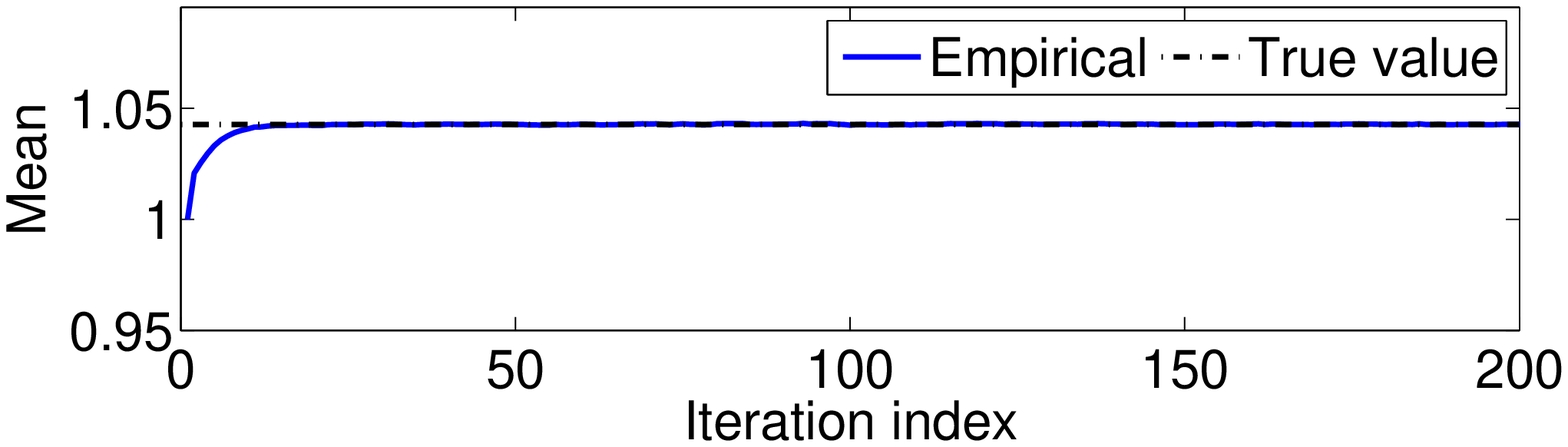}
    \label{fig:mean_est_4nodes_spe_P_skew}
    }
  \subfigure[Variance]{
  \includegraphics[scale = 0.5]{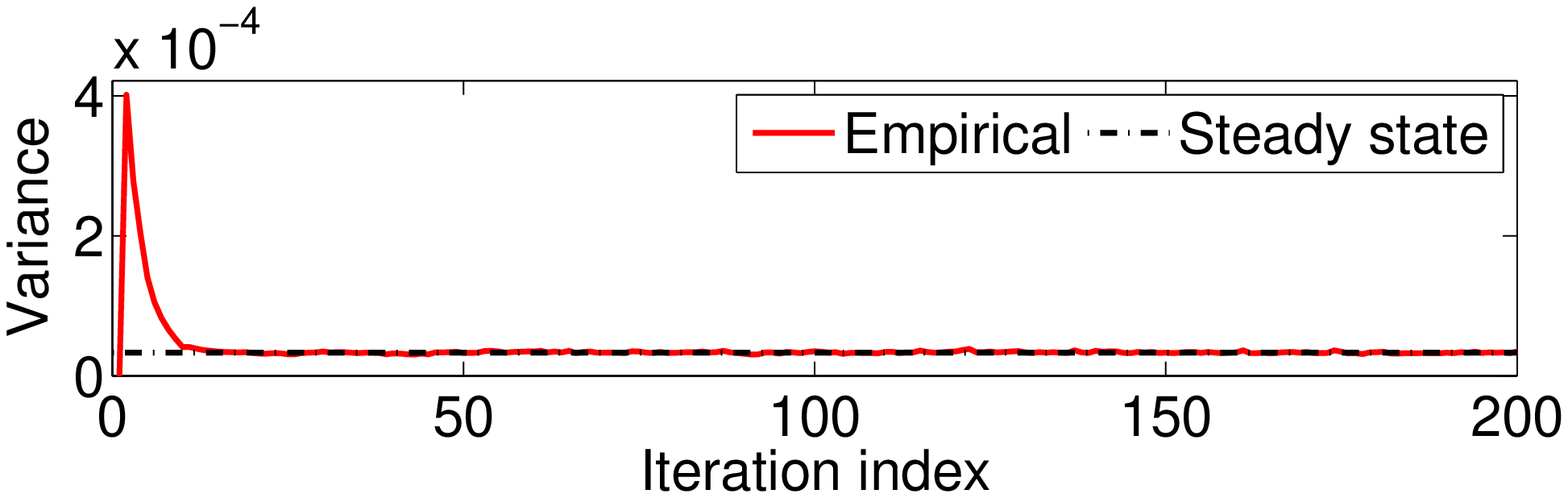}
    \label{fig:var_est_4nodes_spe_P_skew}
    }
    \caption{Mean and variance of the estimate of node $3$'s node
      variable as a function of time. The empirical estimate of mean
      and variance is computed from $1000$ Monte
  Carlo experiments. In $(b)$, the ``steady-state'' corresponds to the
  limiting standard deviation predicted by Lemma~\ref{lem:ss-var}.}
\end{figure}	
Figure~\ref{fig:mean_est_4nodes_spe_P_skew} and Figure~\ref{fig:var_est_4nodes_spe_P_skew} show the empirically
 estimated mean and variance of node $3$'s estimate of its node variable. As predicted by
 Theorem~\ref{thm:ms-convergence-timevarying}, the mean of the
 estimate converges to the true value, since the measurement noise is
 $0$ mean. The variance also converges to the theoretical steady state variance as predicted by Lemma~\ref{lem:ss-var}. 

\subsection{A 100-node network with RWP mobility model}\label{sec:simulations_100nodes}
\begin{figure}[t]
\begin{center}
\includegraphics[scale = 0.5]{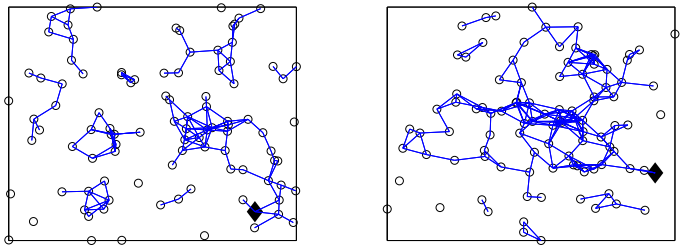}
\caption{Two graphs that occur during a simulation with
  $100$ nodes moving according to the random waypoint mobility model.}\label{fig:two_graphs}
\end{center}
\end{figure}

\begin{figure}[t]
\begin{center}
\psfrag{history}{$\hat{x}_u(k)$}
\psfrag{iteration index}{iteration index $k$}
\includegraphics[scale = 0.5]{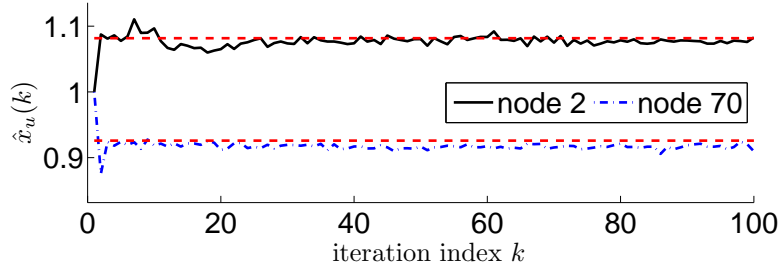}
\caption{The estimates of two nodes in one of the numerical
  experiments involving the $100$-node mobile network.}\label{fig:fg_one_exp_skew}
\end{center}
\end{figure}

\begin{figure}[t]
\begin{center}
\includegraphics[scale = 0.5]{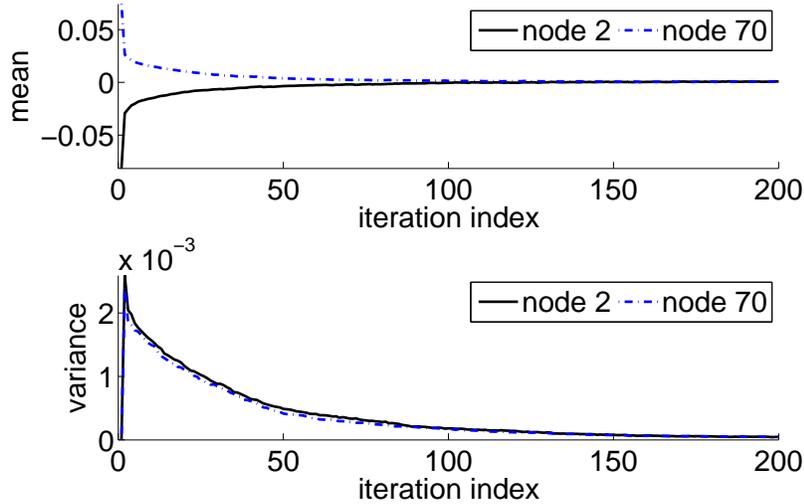}
\caption{Empirically estimated mean and variance of the estimation
  error for one of the nodes in the 100-node mobile network.}\label{fig:fg_mean_var_skew_error}
\end{center}
\end{figure}

Here $100$ nodes move in a $1000\; m \times 1000 \; m$ square
according to the widely used Random Waypoint (RWP) mobility
model~\cite{TC_JB_VD_WCMC:02}. It has been justified in the Appendix A that the graph switching process in this mobility model can be reasonably modeled as a (first order) Markov chain. The parameters maximum/minimum speed and pause time are $v_{min}=10~m/s$, $v_{max}=50~m/s$,
and $t_p=0.1 s$. The communication range is chosen as $100\;m$, and a link failure probability of $0.1$ is used. The mean and
variance of the measurement noise are chosen as $0$ and
$10^{-4}$. Figure~\ref{fig:two_graphs} shows two snapshots of the
network during one of the
simulations. Figure~\ref{fig:fg_one_exp_skew} shows the time trace of
the estimates of two nodes in one of the simulations. The mean and
variance of the estimation error was empirically computed from $1000$
Monte Carlo simulations.
Figure~\ref{fig:fg_mean_var_skew_error} shows mean and variance of the
estimation error for two nodes.  The figure suggests that the
estimates of the node variables converge in the mean square sense.  Note that the transition probability
matrix is not known and the large state space makes it infeasible to
compute the theoretical predictions of limiting mean and variances
that are given in Lemma~\ref{lem:ss-var}. One purpose of these
simulations is therefore to test the performance of the algorithm when
theoretical predictions are not available.

\section{Summary}
We analyzed a distributed algorithm for estimation of clock skew and
offset of the nodes of a mobile network and examined its convergence
properties. The algorithm allows nodes to put different weights on
estimates received from distinct neighbors, depending on the accuracy
of the corresponding relative measurements. The time variation of the
network was modeled as a Markov chain, which makes the algorithm a
jump linear system. Under the assumptions that the Markov chain is
ergodic and the diagonal entries of its transition probability matrix
are positive, the estimates were shown to be mean square convergent as
long as the union of the graphs over time is connected.

Expressions for the asymptotic mean and correlation are also provided
by using results from jump linear systems from~\cite{CostaFragosoMarques:04}. Evaluating
these expressions requires summation of $N$ terms, where $N$ is the
number of distinct graphs that can occur.  In general $N$ is a very
large number, so the utility of these expressions is limited in the
general setting. For instance, if no restriction is placed on the
motion of the nodes or edge formation, $N$ is the number of distinct
graphs possible with $n$ nodes, which is
$2^{\frac{1}{2}n(n-1)}$. Clearly, this is a
very large number unless $n$ is extremely small. However,
in special situations $N$ can be smaller, e.g., if certain nodes are
restricted to move only within certain geographic areas.

In time-varying systems, the rate of change is an important
parameter. The assumption that  Markov chain satisfies
$p_{ii}>0$ provides an upper bound on how fast nodes can move and the
network can change (compared to the time required to obtain relative measurements and
current estimates). This assumption was used to prove
Theorem~\ref{thm:ms-convergence-timevarying}. However, it is possible
that mean square convergence can be proved with weaker constraints on
the speed of topology change.
 
We have not examined the question of convergence rate. It is likely that the transition
probabilities of the chain will play a role in the convergence
rate. However, precisely characterizing of the convergence rate of the
algorithm remains an open problem. The time to reach acceptable
estimation accuracy can however be reduced by more careful choice of
the initial condition, e.g., using the flagged initialization scheme
proposed in~\cite{PB_NdS_JH_LNCS:06}. 

\bibliographystyle{IEEEtran}
%PB
\bibliography{../../../PBbib/Barooah,../../../PBbib/sensnet_bib_dbase,../../../PBbib/CLBib}
%CL
%\bibliography{../Pbbib/Barooah,../PBbib/sensnet_bib_dbase,../PBbib/CLBib}

%\bibliographystyle{plain}        % Include this if you use bibtex 
%\bibliography{autosam}           % and a bib file to produce the 
                                 % bibliography (preferred). The
                                 % correct style is generated by
                                 % Elsevier at the time of printing.

%\begin{thebibliography}{99}     % Otherwise use the  
                                 % thebibliography environment.
                                 % Insert the full references here.
                                 % See a recent issue of Automatica 
                                 % for the style.
%  \bibitem[Heritage, 1992]{Heritage:92}
%     (1992) {\it The American Heritage. 
%     Dictionary of the American Language.}
%     Houghton Mifflin Company.
%  \bibitem[Able, 1956]{Abl:56}
%     B.~C.~Able (1956). Nucleic acid content of macroscope. 
%     {\it Nature 2}, 7--9. 
%  \bibitem[Able {\em et al.}, 1954]{AbTaRu:54}   
%     B.~C. Able, R.~A. Tagg, and M.~Rush (1954).
%     Enzyme-catalyzed cellular transanimations.
%     In A.~F.~Round, editor, 
%     {\it Advances in Enzymology Vol. 2} (125--247). 
%     New York, Academic Press.
%  \bibitem[R.~Keohane, 1958]{Keo:58}
%     R.~Keohane (1958).
%     {\it Power and Interdependence: 
%     World Politics in Transition.}
%     Boston, Little, Brown \& Co.
%  \bibitem[Powers, 1985]{Pow:85}
%     T.~Powers (1985).
%     Is there a way out?
%     {\it Harpers, June 1985}, 35--47.

%\end{thebibliography}

\ifthenelse{\equal{\PaperORReport}{Paper}}{}{
\appendix
\section{Markovian model of topology change}\label{sec:markov_modeling}
Here we examine the question of the applicability of the Markovian
model of graph switching. An example in which the time variation of the graphs satisfies the
homogeneous Markov model is a network of mobile agents whose motion is
modeled with first order dynamics with range-determined
communication. In ad-hoc networks literature this is referred to as
the \emph{random walk mobility model}~\cite{TC_JB_VD_WCMC:02}. Specifically, suppose the position of node $u$ at time
$k$, denoted by $p_u(k)$, is restricted to lie on the unit sphere
$\mathbf{S}^2 = \{ x \in \R^3| \|x\|=1\}$, and suppose the position
evolution obeys: $p_u(k+1) = f(p_u(k) + \Delta_u(k))$, where
$\Delta_u(k)$ is a stationary zero-mean white noise sequence for every
$u$, and $\Exp[\Delta_u(k) \Delta_v(k)^T] = 0$ unless $u=v$. The
function $f(\cdot): \R^3 \to \mathbf{S}^2$ is a projection function
onto the unit-sphere. In addition, $(u,v) \in \E(k)$ if and only if
the geodesic distance between them is less than or equal to some
predetermined value. In this case, the graph $\G(k)$ is uniquely
determined by the node positions at time $k$, and the prediction of $\G(k+1)$ given
$\G(k)$ cannot be improved by the knowledge of the graphs observed
prior to $k$: $\G(k-1),\dots,\G(0)$. Hence the evolution of the graph
sequence satisfies the Markovian property. If in addition random
communication failure leads to two nodes not being able to communicate
even when they are in range, the Markovian property is retained if the
communication failure is i.i.d. 

However, it is not straightforward to
check if the sequence of graphs generated by the model satisfies the
Markovian property for other mobility models. A general method of checking Markovian switching
of graphs is therefore needed. We borrow a method that is proposed in~\cite{CC:73} to check if a
stochastic process is Markov from observations of the process. We
first introduce some standard notation from information theory. Let
$X$ be a discrete random variable with sample space
$\Omega=\{1,\dots,N\}$ and probability mass function $p(x)= \Prob(X=x)$,
where $x\in\Omega$. The entropy of $X$ is defined by
\begin{align}
H(X)=-\sum_{x\in\Omega} p(x)\log\ p(x). 
\end{align} 
The definition of entropy is extended to a pair of random variable
$X,Y$, where $X,Y\in\Omega$, as follows
\begin{align}
H(X,Y) \eqdef -\sum_{x,y\in\Omega} p(x,y)\log\ p(x,y). 
\end{align} 
The conditional entropy $H(Y|X)$ is defined as
\begin{align}
H(Y|X) \eqdef -\sum_{x,y\in\Omega} p(x,y)\log\ p(x|y)=H(X,Y)-H(X),
\end{align}
where $p(x,y)$ is joint probability mass function. The conditional entropy measures the conditional uncertainty about an
event given the another event. Consider a stochastic process $\{X_1,
X_2,\dots\}$. Assuming the process is stationary,  we denote
$H(single)\eqdef H(X_k)$ and $H(double)\eqdef H(X_{k},X_{k+1})$, for
all $k=1,\dots$. It is straightforward to show that
$H(double)=2H(single)$ if the successive random variables $X_k$ are
i.i.d. In this case the random process is a zero-order Markov
process. If the random variables $X_k$ are not
independent, $H<H(double)<2H$. To address the question of whether it is
($m$-th order) Markov, we extend the entropy definition
to multivariate random variables, with $H(triple)\eqdef
H(X_{k},X_{k+1},X_{k+2})$, etc. Now, the sequence
$H_0=\log(N)$, $H_1=H(single)$, $H_2=H(double)-H(single)$ and
$H_3=H(triple)-H(double)$, etc, measures the conditional uncertainty
for each order of dependence. A graphical approach is given
in~\cite{CC:73} to determine the order of dependence of a random
process by plotting the estimates of each $H_i$, where $i=1,2,\dots$
and examining the shape of the curve. 
\begin{figure}
\centering 
\psfrag{H}{$\hat{H}_i$}
 \subfigure[Independent]{
  \includegraphics[scale = 0.2]{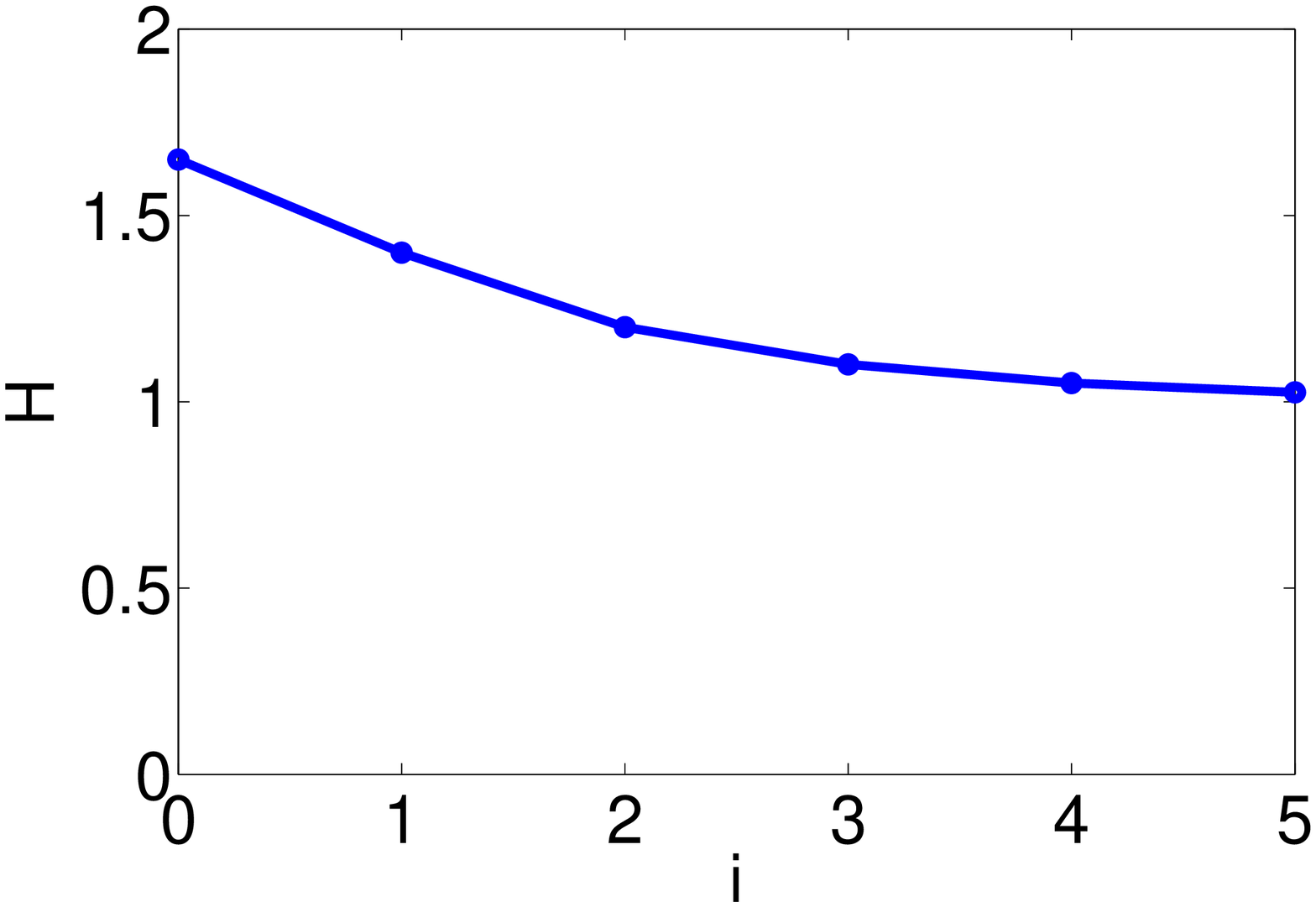}
    \label{fig:markov_check_0}
    }
  \subfigure[First-order]{
  \includegraphics[scale = 0.2]{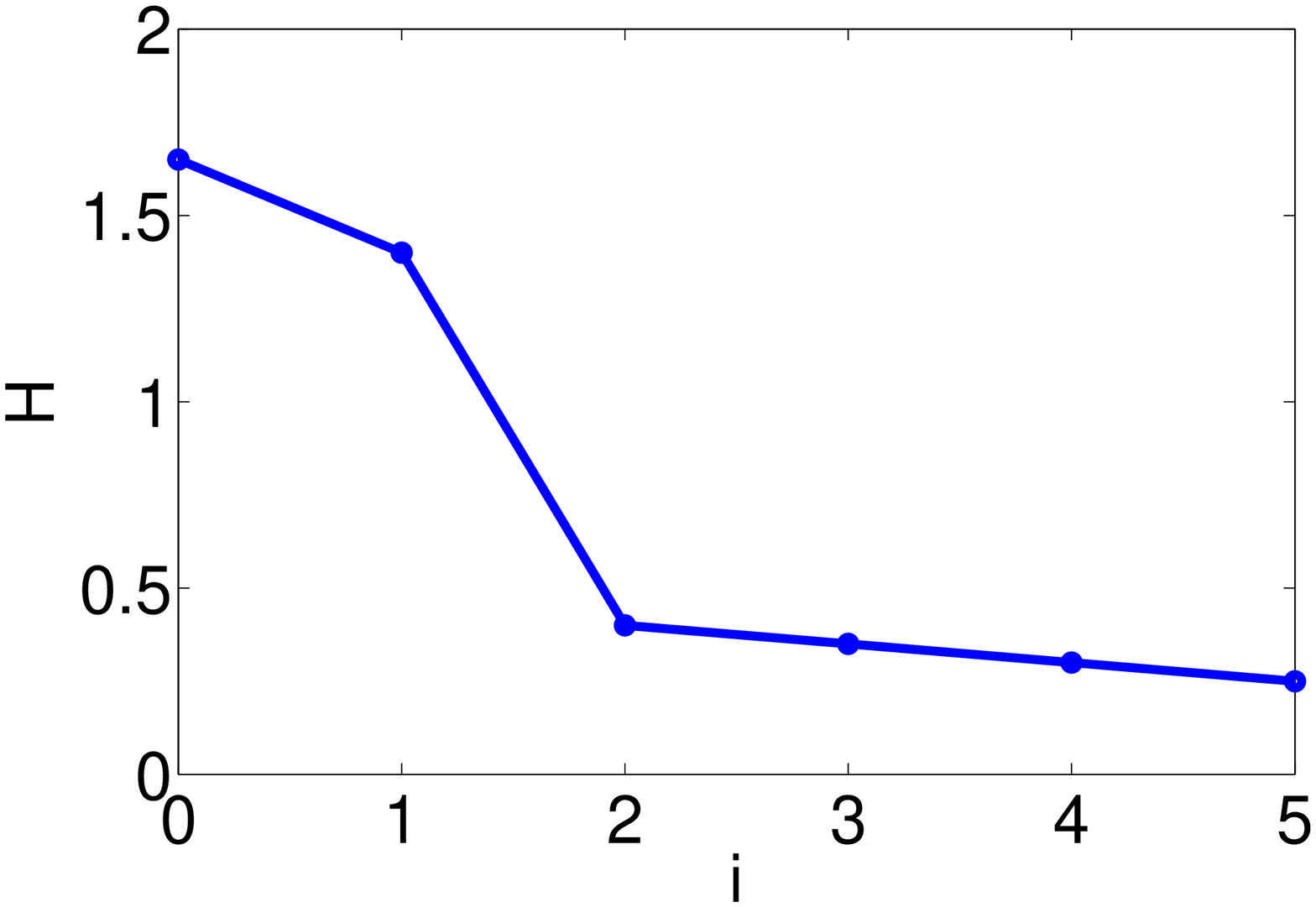}
    \label{fig:markov_check_1}
    }
  \subfigure[Second-order]{
  \includegraphics[scale = 0.2]{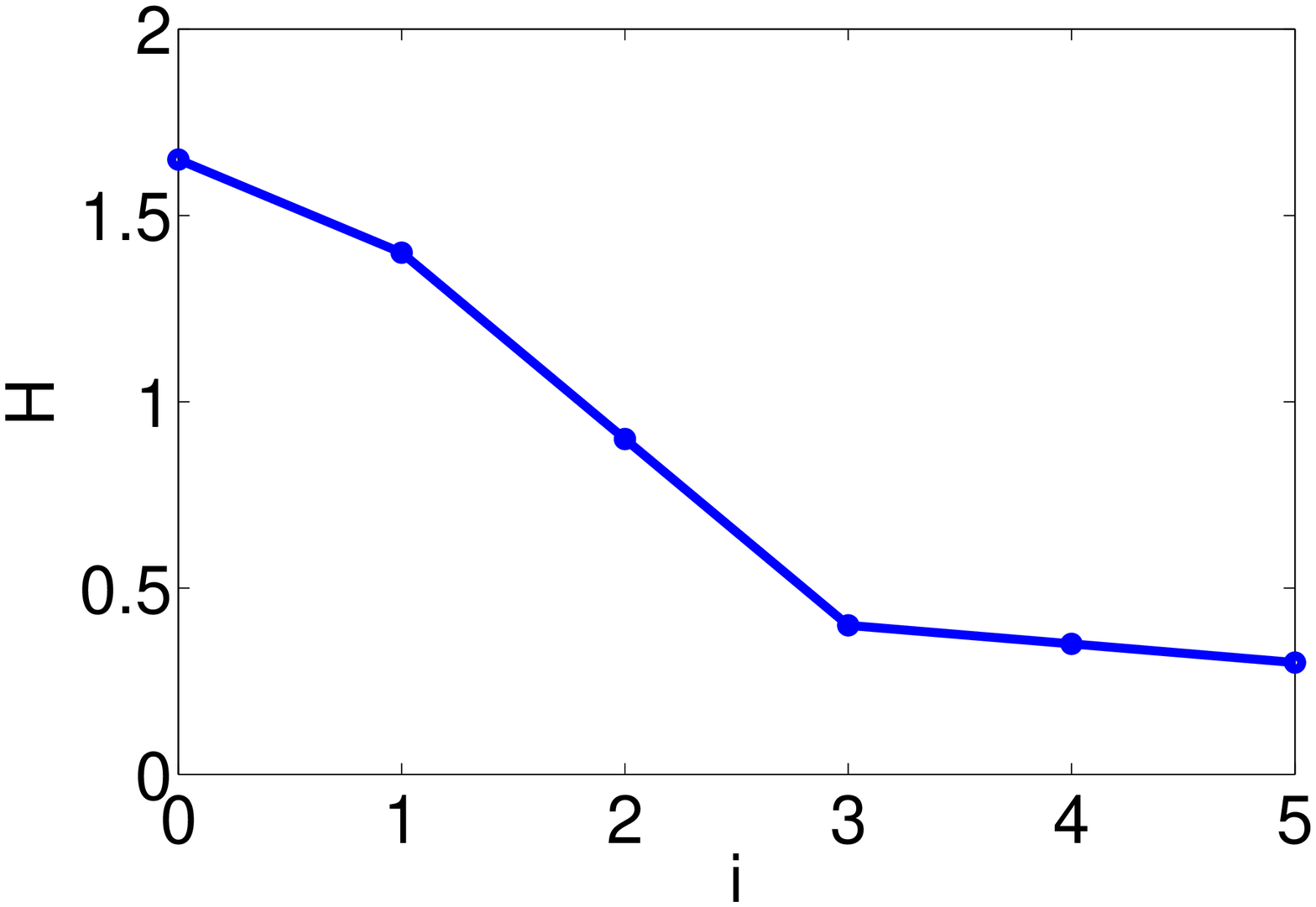}
    \label{fig:markov_check_2}
     }
    \caption{The standard shapes of estimated conditional entropy for
      three different cases: \subref{fig:markov_check_0} independence,
      \subref{fig:markov_check_1} first-order dependence and
      \subref{fig:markov_check_2} second-order dependence. If a
      process is a first order Markov chain, empirically computed
      conditional entropy will show a trend similar to the one in (b).}\label{fig:markov_check_all}
\end{figure} 
The estimate of each $H_i$ can be calculated from
observations. Figure~\ref{fig:markov_check_all} shows the standard
shapes for independence, first-order dependency, and second-order
dependency. If the process is independent, then knowing the value of
$X_k$ will not help in predicting $X_{k+1}$, which is seen in the flat
shape of the entropy function in Figure~\ref{fig:markov_check_0}. In
contrast, the sharp drop from $\hat{H}_1$ to $\hat{H}_2$ in
Figure~\ref{fig:markov_check_1} indicates that knowing the value of
$X_{k-1}$ will dramatically decrease the uncertainty in the prediction
of $X_k$, while the values of $X_{k-2}, X_{k-3}, \dots$ will not help
much. This accords with the dependence property of a first-order
Markov chain. Similarly, Figure~\ref{fig:markov_check_2} indicates
that the previous two variables $X_{k-2},X_{k-1}$ are both important to
predict $X_k$. In this case the process is better modeled as a second
order Markov chain.

In order to conclude whether the evolution of graphs is governed by a
first-order Markov chain, we adopt the method discussed above as
follows. For a particular mobility model, we conduct a simulation and
collect observations of the graph sequence. Since the underlying
sample space $\graphset$ of the stochastic
process $\G(k)$ is finite, the method described above is applicable. We then use the approach
above to check whether the plot of $\hat{H}_i$ estimated from the
collected observations is closer to that in
Figure~\ref{fig:markov_check_1} than to those in
Figure~\ref{fig:markov_check_0} or Figure~\ref{fig:markov_check_2}. If
so, we declare that it is \emph{reasonable to model the graph
  switching process as Markovian}.

%% application of the method ... 
As an illustrative example, we consider the widely used \emph{random
waypoint} (RWP) mobility model~\cite{TC_JB_VD_WCMC:02}. In the RWP
model, each node is initialized to stay in its initial position for a
certain period of time (so called pause time $t_p$). Then, the node
picks a random destination within the region it is allowed to move and
a speed that is uniformly distributed in $[v_{min}, v_{max}]$. Once
node reaches the new destination, it pauses again for $t_p$ before
starting over. We conduct a simulation of the RWP model with 3 nodes,
where $v_{min}$, $v_{max}$, $t_p$ are chosen as $10\;m/s$, $50\;m/s$
and $0.1\;s$. The nodes are allowed to move in a region $10 \times 10$
m. Nodes' positions are initialized randomly.  The sample space consists of $8$
graphs.  By performing the
simulation for a long time ($10^4\;s$), we obtain a large number of observations of the process $\{\G(k)\}$. The probability mass function is empirically estimated
from the observations. For estimating conditional entropies, certain
conditional probabilities, especially those of the type
$P(\G(k)= G_1|\G(k-1)=G_2,\G(k-2)=G_3)$, are problematic since the
relevant events may not be observed even in a very long sequence of
observations. In this case we set the corresponding probabilities to
$0$ and use $0 \log 0 = 0$. The empirically estimated
conditional entropies $\hat{H}_i$ are shown in
Figure~\ref{fig:markov_check}. Clearly, the shape of curve is similar
to that in Figure~\ref{fig:markov_check_1}. Therefore, we conclude that the graph switching process in RWP mobility can be reasonably modeled as a (first
order) Markov chain. 
\begin{figure}[t]
\begin{center}
\psfrag{H}{$\hat{H}_i$}
\includegraphics[scale = 0.25]{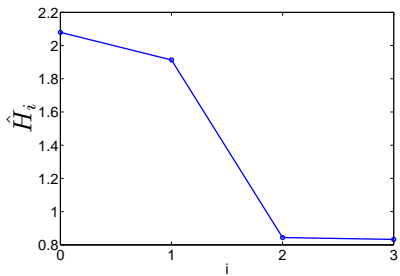}
\caption{Empirically estimated conditional entropy for the graph
  process $\{\G_k\}$ with three nodes moving according to the random
  waypoint mobility model.}\label{fig:markov_check}
\end{center}
\end{figure}

Note that in RWP mobility, prediction of the future node locations (and
therefore the graph) based on knowledge of past and present may be
more accurate instead of prediction based on only the
present. Therefore it is quite possible that the graph switching is
not first-order Markov. However, the results of the test above shows
that a Markov model quite accurately captures the graph switching
process with RWP mobility.

\section{Proof of Lemma~\ref{lemma:rho_of_D}}\label{proof-lemma:rho_of_D}
Recall that a non-negative matrix is
called stochastic matrix if each row sum is $1$. If $X$ is a stochastic
matrix, then $\rho(X)=1$~\cite{Minc:88}.
%%%%%%%%%%%%%%%%%%%%%%%
\begin{prop}\label{prop:JLS-MSS-spetral-big_matrix}
 \begin{enumerate}
  \item If $X$ is stochastic matrix, $X \otimes X$ is also a
    stochastic matrix.
  \item The matrices $J_i$ and $F_i$,  $i=1,\dots,N$,
    defined in~\eqref{eq:def_J_bar} are stochastic matrices.
\item Let
\begin{align*}
  K & \eqdef \left(\begin{array}{cccc}
       F_1& F_2 &\cdots&  F_N \\
       F_1&F_2&\cdots&  F_N \\
       \vdots & \vdots & \ddots & \vdots \\
       F_1& F_2&\cdots&   F_N    \\ 
\end{array}\right)_{Nn^2\times Nn^2}, & K_b & \eqdef \left(\begin{array}{cccc}
       F_{b1}& F_{b2} &\cdots&  F_{bN} \\
       F_{b1}&F_2&\cdots&  F_{bN} \\
       \vdots & \vdots & \ddots & \vdots \\
       F_{b1} & F_{b2}&\cdots&   F_{bN}    \\ 
\end{array}\right)_{Nn_b^2 \times Nn_b^2},
\end{align*}
There exists a permutation matrix $X$ so that $K_b$ is a principal sub-matrix of $  X^{T} K X$.  
\item For the matrix $\mathcal{D}$ defined in~\eqref{eq:D-bar-def}, $\rho(\mathcal{D})\leq 1$. 
  \end{enumerate}
\end{prop}
%%%%%%%%%%%%
\begin{pf*}{Proof:}
The first two statements are straightforward to establish. The third
statement follows from the fact that $J_{bi}$ is a principal submatrix of $J_i$.
We therefore prove only the fourth
statement. From~\eqref{eq:define_D}, $\rho(\mathcal{D})
=\rho([p_{ji}F_j])$. Since $p_{ji}F_j$ is a non-negative square matrix, it follows
from~\cite[Theorem 3.2]{MQC_XZL:04} that $\rho([p_{ji}F_j]) \leq
\rho([\|p_{ji}F_j \|_\infty])$. Moreover, $\|p_{ji}F_j
\|_\infty = p_{ji}\|F_j \|_\infty$. Since $F_j$ is a
stochastic matrix, $\|F_j\|_\infty=1$. We therefore have
 \begin{align*}  
      \rho(\mathcal{D}) \leq \rho([p_{ji}\|F_j \|_\infty])
     = \rho([p_{ji}])=\rho(\MarkovP^T)=\rho(\MarkovP) =1. \tag*{\qed}
\end{align*}
\end{pf*}

\medskip

\begin{lem}{\label{lemma:JLS-MSS-irreducible-big-matrix}}
  Let $\MarkovP$ be the transition probability matrix of an $N$-state
  ergodic Markov chain whose diagonal entries are positive. The
  $\mathcal{D}$ defined in~\eqref{eq:define_D} is irreducible if and
  only if the union graph $\hat{\G}$ defined in~\eqref{eq:G-union} is
  connected. 
\end{lem}

\medskip
The proof of Lemma~\ref{lemma:JLS-MSS-irreducible-big-matrix} is postponed and we first prove Lemma~\ref{lemma:rho_of_D}.

\begin{pf*}{Proof of Lemma~\ref{lemma:rho_of_D}}
  Since the union graph $\hat{\G}$ is connected, it follows from
  Lemma~\ref{lemma:JLS-MSS-irreducible-big-matrix} that $\mathcal{D}$
  is irreducible. From the third statement of Proposition~\ref{prop:JLS-MSS-spetral-big_matrix},
  there exists a permutation matrix $X$, such that $\mathcal{D}_b$ is
  a principal submatrix of $X^{T}\mathcal{D}X$. The spectral radius
  of an irreducible matrix is strictly greater than the spectral
  radius of any of its principal submatrices, which follows from
  Theorem 5.1 in~\cite{Minc:88}. Therefore we have
\begin{align*}
\rho(\mathcal{D}_b)<\rho(X^{T}\mathcal{D}X).  
\end{align*}
From the fourth statement in Proposition~\ref{prop:JLS-MSS-spetral-big_matrix} and the fact that
permutation does not change eigenvalues, it follows that
\begin{align*}
\rho(X^{T}\mathcal{D}X)=\rho(\mathcal{D})\leq 1.  
\end{align*}
Combining these two inequalities we get that if $\hat{\G}$ is connected
then  $\rho(\mathcal{D}_b) <1$. To prove necessity, we construct a
counterexample, in particular, a trivial Markov chain with a single
state: $\graphset = \{\G_1\}$ (so that $\MarkovP = 1$) where $\G_1$
is a $n$-node graph without any connected edge.  Then $\mathcal{D}_b = F_{b1}
= J_{b1} \otimes J_{b1} = I$, which has a spectral radius of
unity. This completes the proof of the lemma. \qed
\end{pf*}

The proof of Lemma~\ref{lemma:JLS-MSS-irreducible-big-matrix} needs the following definition and results.
All matrices are non-negative hereafter; so we will explicitly say
``non-negative'' only when we have to stress it. For matrices
$X_1,X_2$ of same dimension, we say $X_1$ and $X_2$ are
\emph{congruent}, and write $X_1\cong X_2$, if the following holds:
${(X_1)}_{\imath\jmath}\neq 0$ if and only if
${(X_2)}_{\imath\jmath}\neq 0$. We also write $X_1\succeq X_2$ if the
following condition is satisfied: ${(X_1)}_{\imath\jmath}\neq 0$ if
${(X_2)}_{\imath\jmath}\neq 0$. The directed graph
$\vec{G}(X)=(\V(X),\vec{\E}(X))$ corresponding to a square matrix $X
\in \R^{n\times n}$ is a graph defined on $n$ nodes in which
$(u,v)\in\vec{\E}(X)$ if and only if $X_{u,v} \neq 0$.  A directed
graph $\vec{G}$ is called \emph{strongly connected} if for each pair
of nodes $u$ and $v$, there is a sequence of directed edges in
$\vec{\E}$ leading from $u$ to $v$~\cite{CarlMeyer:01}. If $\vec{G}_1$
is a subgraph of $\vec{G}_2$, meaning that $\vec{G}_2$ contains all
the nodes and edges of $\vec{G}_1$, we write $\vec{G}_1 \subseteq
\vec{G}_2$ or $\vec{G}_2\supseteq \vec{G}_1$.  Two directed graphs
$\vec{G}_1$ and $\vec{G}_2$ are called \emph{congruent} if their
adjacency matrices are congruent. We denote by $Adj(\vec{G})$ the
adjacency matrix of the graph $\vec{G}$. For a $n \times n $ square matrix $X$, we
write $Adj(X)$ to denote $Adj(\vec{G}(X))$, which is an $n \times n$
matrix with $\imath,\jmath$-th entry equal to $1$ if and only if
$X_{\imath\jmath} > 0$, and $0$ otherwise. Essentially, the matrix
$Adj(X)$ replaces the positive entries of $X$ by $1$ and leaves the
$0$ entries untouched.

The following statements for non-negative matrices can be verified in
a straightforward manner.  All the matrices are of the same dimension.
\begin{prop}{\label{prop:JLS-MSS-cartesian-property}}
\begin{enumerate}
\item $X \cong Adj(X)$.
\item $\vec{G}(X_1)\cong \vec{G}(X_2)$ if and only if $X_1\cong X_2$. 
\item $\vec{G}(X_1)\supseteq\vec{G}(X_2)$ if $X_1\succeq X_2$. 
\item $\vec{G}(\sum_i^\ell X_i)\cong \cup_{i=1}^{\ell}\vec{G}(X_i)$.

\end{enumerate} 
\end{prop}
%If  $X$ is symmetric,\PBcomment{probably unnecessary}, $\vec{G}(X)$ can be simplified to undirected graph $G(X)=(\V(X),\E(X))$ and so does the relevant notations.
\begin{prop}{\label{prop:JLS-MSS-irreducible-property}}
The graph $\vec{G}(X)$ is strongly connected if and only if $X$ is
irreducible. If $\vec{G}(X)$ is strongly
connected, then $\vec{G}(X \otimes X)$ is also strongly connected and
thus $X \otimes X$ is irreducible.
\end{prop}
The first statement of the proposition is
well-known~\cite[pp.671]{CarlMeyer:01}. The second statement follows
from the first in a straightforward manner. 

Now we define the \emph{Cartesian product} of two directed graphs $\vec{G}_1=(\V_1,\vec{\E}_1)$ and
$\vec{G}_2=(\V_2,\vec{\E}_2)$, which is denoted by $\vec{G}_1 \Box
\vec{G}_2$. The Cartesian product has the vertex
set equal to $\V_1 \times \V_2$, so that nodes in the product are
denoted by the pair $(u,v)$, with $u \in \V_1$ and $v \in \V_2$, which
is not to be confused with an edge. \emph{In order to prevent confusion, we
will denote an edge from $u$ to $v$ in the sequel by $u \rightarrow v$}. The edge set of the Cartesian product is
characterized by the following property: there is an edge
$(u_1,v_1)\rightarrow (u_2,v_2)$ in $\vec{G}_1 \Box \vec{G}_2$ if
either $u_1=u_2$ and $v_1 \rightarrow v_2 \in
\vec{\E}_2$ or $v_1=v_2$ and $u_1 \rightarrow u_2 \in \vec{E}_1$. Cartesian
products of undirected graphs $\G_1$ and $\G_2$ are similarly defined, except
that the resulting product graph is also undirected.
The following properties will be useful in future.
\begin{prop}\label{prop:Cartesian-graphs}
\begin{enumerate}
\item If $\vec{G}_1$ and $\vec{G}_2$ are strongly connected, so is $\vec{G}_1\Box \vec{G}_2$. 
\item If $Adj(X_1)$ and $Adj(X_2)$ are symmetric, then,
\begin{align}\label{eq:Cartesian-graph-adj}
  Adj(\vec{G}(X_1)\Box \vec{G}(X_2))=Adj(X_1)\otimes I +I\otimes Adj(X_2).
\end{align}
%\item If $Adj(X_1)$ and $Adj(X_2)$ are symmetric and irreducible,  $\vec{G}(X_1)\Box \vec{G}(X_2)$ is strongly connected. 
\end{enumerate}  
\end{prop}
\begin{pf*}{Proof of Proposition~\ref{prop:Cartesian-graphs}} The first
  statement is from~\cite[Table 2]{FH_CT:66}.  To prove the second statement, we introduce the notation $G(X)$, which is the undirected graph corresponding to a symmetric matrix $X$. It follows that $Adj(\vec{G}(X_1)\Box \vec{G}(X_2))=Adj(G(Adj(X_1))\Box G(Adj(X_2)))$. From~\cite[Section 2.3]{BY_EC_TL:09},  $Adj(G(Adj(X_1))\Box G(Adj(X_2)))=Adj(X_1)\otimes I +I\otimes Adj(X_2)$. Thus, we prove \eqref{eq:Cartesian-graph-adj}. \qed
%  To prove the 
%second statement, note that $Adj(X)$ is
%symmetric and irreducible, it is the adjacency matrix of a connected
%undirected graph\PBcomment{good to have a citation here. Chenda, does
%  Godsil and royle have something about this?}. Therefore, if
%$Adj(X_1)$ and $Adj(X_2)$ are symmetric and irreducible, the Cartesian
%product of the corresponding undirected graphs $G(X_1)$ and $G(X_2)$
%is a connected undirected graph. The result now follows from the property that Cartesian product of
%two undirected graphs is connected if and only if both graphs are
%connected~\cite[Proposition 1.34]{WI_SK:00}, and that $\vec{\G}(X)$ is
%strongly connected if $X$ corresponds to a connected undirected
%graph. \PBcomment{need some more work - undicted graph $G(X)$ was not
%  defined but used here. Are we also assuming the reader knows basic
%  things about graphs? We should define/discuss everything that is needed.}
\end{pf*}

The following results will be useful in the proof of Lemma~\ref{lemma:JLS-MSS-irreducible-big-matrix}.
\begin{prop}\label{prop:union_graph_connected}
If the union graph $\hat{\G} = \cup_{i=1}^{N}\G_i$ is connected, then
$\cup_{i=1}^{N}\vec{G}(F_i)$ is strongly connected.
\end{prop}
\begin{pf*}{Proof of Proposition~\ref{prop:union_graph_connected}}
Recall that $F_i =J_i\otimes J_i$, where $J_i
=(M_i+D_i)^{-1}(N_i+D_i)$. Since $M_i+D_i$, and therefore its inverse, is a diagonal matrix with
positive entries, $J_i \cong (N_i + D_i)$. By property 2 of Proposition~\ref{prop:JLS-MSS-cartesian-property},
\begin{align}
\cup_{i=1}^{N}\vec{G}(F_i) &\cong \vec{G}(\sum_{i=1}^NF_i) \cong \vec{G}\left(\sum_{i=1}^N\{(N_i+D_i)\otimes(N_i+D_i)\}\right)
\end{align}
We also have 
\begin{align}
(N_i+D_i)\otimes(N_i+D_i) & \succeq D_i\otimes N_i + N_i\otimes D_i
\end{align}
by dropping two terms in the expansion using their
non-negativity. Since $D_i \cong I$, we get
\begin{align*}
(N_i+D_i) \otimes (N_i+D_i) & \succeq I \otimes N_i + N_i\otimes I \\
\Rightarrow \sum_{i=1}^N \left( (N_i+D_i) \otimes (N_i+D_i)\right) &
\succeq  I \otimes \left( \sum_{i=1}^N N_i \right) + \left( \sum_{i=1}^N N_i \right) \otimes I. 
\end{align*}
Using property 3 in Proposition~\ref{prop:JLS-MSS-cartesian-property},
we have
\begin{align}\label{eq:Gunion-Fis}
\cup_{i=1}^{N}\vec{G}(F_i) & \supseteq \vec{G} \left((\sum_{i=1}^N N_i)\otimes I+ I\otimes (\sum_{i=1}^N N_i) \right) 
 \cong \vec{G}( \sum_{i=1}^N N_i) \Box \vec{G}(\sum_{i=1}^N N_i) .
\end{align}
where the congruence follows from the second statement in
Proposition~\ref{prop:Cartesian-graphs}. Recall the structure of $N_i$, it follows that (i) $Adj(\G_i) =
Adj(N_i)$, and (ii)$Adj(N_i)$ is
symmetric. As a result, $Adj(\sum_{i=1}^N N_i)$ is also
symmetric. Since $\hat{\G}=\cup_{i=1}^{N}\G_i$ is a connected
undirected graph, its adjacency matrix is irreducible. This means
$Adj(\cup_{i=1}^{N}\G_i) = \sum_{i=1}^N Adj(\G_i) = \sum_{i=1}^N
Adj(N_i)$ is irreducible. Due to the first statement in
Proposition~\ref{prop:Cartesian-graphs}, $\vec{G}( \sum_{i=1}^N N_i)
\Box \vec{G}(\sum_{i=1}^N N_i) $ is strongly connected. The result of this proposition now
follows from~\eqref{eq:Gunion-Fis}. \qed
\end{pf*}

The \emph{Kronecker product} of two graphs
$\vec{G}_1=(\V_1,\vec{\E}_1)$ and $\vec{G}_2=(\V_2,\vec{\E}_2)$,
denoted by $\vec{G}_1\otimes \vec{G}_2$, has the vertex set equal to
$\vec{V}_1 \times \vec{V}_2$ and an edge set that is characterized by
the following property: there is an edge $(u_1,v_1)\rightarrow (u_2,v_2)$
in $\vec{G}_1\otimes \vec{G}_2$ if and only if $u_1 \rightarrow u_2\in
\vec{\E}_1$ and $v_1 \rightarrow v_2\in
\vec{\E}_2$ \cite{FH_CT:66}. Note that the Cartesian and Kronecker products
$\vec{G}_1 \Box \vec{G}_2$ and $\vec{G}_1 \otimes \vec{G}_2$ have
the same vertex sets but distinct edge sets. We have the
following property of Kronecker product of graphs from \cite{PW_62}:
\begin{align}\label{eq:Kronecker-graph-adj}
Adj(\vec{G}_1\otimes \vec{G}_2)=Adj(\vec{G}_1)\otimes Adj(\vec{G}_2).  
\end{align}
Adjacency matrices of both Cartesian and Kronecker products of two graphs
are related to the adjacency matrices of the individual graphs through
the matrix Kronecker product, cf.~\eqref{eq:Cartesian-graph-adj} and~\eqref{eq:Kronecker-graph-adj}.

Now we are ready to prove the Lemma~\ref{lemma:JLS-MSS-irreducible-big-matrix}.
%%%%%%%%%%%%%%%%%%%%%%%%%%%%%%%%%%%%%%%%%%%%%%%%%%%%
\begin{pf*}{Proof of Lemma~\ref{lemma:JLS-MSS-irreducible-big-matrix}}
(Connectivity $\Rightarrow$ irreducibility): 
Here we have to prove that if the union graph $\hat{\G}$ is connected
then the matrix $\mathcal{D}$ is irreducible. We will prove it by
showing that the directed graph $\vec{\G}(\mathcal{D})$ is strongly connected.
  Let $Z_j$ and $S_j$ be the diagonal and off-diagonal parts of
  $F_j$. Since $Z_i$ is a  non-negative matrix with positive diagonal,
  we get
\begin{align}
\mathcal{D}  & \cong [p_{ji}Z_j] + [p_{ji}S_j]  \cong [p_{ji}I_{n^2}] + [p_{ji}S_j] \notag\\
                             & \succeq\MarkovP^T\otimes I+ diag[p_{ii}S_i]
\end{align}
where we have used the fact that $[p_{ji}I_{n^2}] = \MarkovP^T
\otimes I$ and dropped the off-diagonal blocks of
$[p_{ji}S_j]$. Therefore
\begin{align*}
\vec{\G}(\mathcal{D}) & \supseteq \vec{G}(\MarkovP^T \otimes I )
\bigcup  \vec{G}(diag[S_i]) \\
& = \{\vec{G}(\MarkovP^T)\otimes \vec{G}(I) \}\bigcup  \vec{G}(diag[S_i])
\end{align*}
where the equality follows from the
property~\eqref{eq:Kronecker-graph-adj} of Kronecker product of
graphs. We will now show that the directed graph
$\{\vec{G}(\MarkovP^T)\otimes \vec{G}(I) \}\bigcup  \vec{G}(diag[S_i])$
is strongly connected, which proves that $\vec{\G}(\mathcal{D})$ is as well.

First notice that there are $Nn^2$ nodes in the graph
$\vec{\G}(\mathcal{D})$, so are $\MarkovP^T\otimes I$ and $diag[S_i]$. It is convenient to imagine them as $N \times N$ block matrices,
with each block being of dimension $n^2 \times n^2$. Therefore, we introduce a useful new notation. Let's index
a node by the pair $(p_i,d_\kappa)$, which is the $((i-1)N+\kappa)$-th node in $\vec{\G}(\mathcal{D})$, where $i=1,\dots,N$ and
$\kappa=1,\dots,n^2$. This notation is similarly suitable for $\MarkovP^T\otimes I$ and $diag[S_i]$. To prove that $\{\vec{G}(\MarkovP^T)\otimes \vec{G}(I) \}\bigcup  \vec{G}(diag[S_i])$
is strongly connected, we need to show the following
\begin{align}\label{eq:lemma-statement}
\text{\emph{There is a path from an arbitrary node}} (p_i,d_\kappa) \text{\emph{to another
  arbitrary node}} (p_j,d_\nu) \notag \\
\text{ \emph{in the graph} } \{\vec{G}(\MarkovP^T)\otimes \vec{G}(I) \}\bigcup  \vec{G}(diag[S_i]).  
\end{align}
 The following properties will be used to construct a proof of~\eqref{eq:lemma-statement}:
\begin{enumerate}
\item {\bf s1:} There exists a path from $(p_{i},d_{\kappa})$ to
  $(p_{j},d_{\kappa})$ in $\vec{G}(\MarkovP^T \otimes I)$ for all
  $i,j=1,\dots,N$ and $\kappa=1,\dots,n^2$. 
\item {\bf s2:} If $d_\kappa \rightarrow d_h$ is an edge in $\vec{\G}(S_\ell)$, then
  $(p_\ell,d_\kappa) \rightarrow (p_\ell,d_h)$ is an edge in $\vec{\G}(diag
  [S_i])$. 
\end{enumerate}
The first statement is proved
  as follows. Since the Markov chain is
  ergodic, $\MarkovP$ - and therefore  $\MarkovP^T$ - is irreducible,
  which means $\vec{\G}(\MarkovP^T)$ is strongly connected. Thus,
  given arbitrary nodes $p$ and $q$ in $\vec{\G}(\MarkovP^T)$,
  there is a path connecting them in $\vec{\G}(\MarkovP^T)$. Call this
  path $p,u_1,u_2,\dots,u_m,q$. Since the edge $d_\kappa \rightarrow 
  d_\kappa \in \vec{\G}(I)$ exists for every $\kappa$, it now follows from the definition of Kronecker product of
  graphs that the path $(p,d_\kappa), (u_1,d_\kappa),(u_2,d_\kappa),\dots,(u_m,d_\kappa),(q,d_\kappa)$ exists
  in $\vec{G}(\MarkovP^T)\otimes\vec{G}(I)$ for every $\kappa =
  1,\dots, n^2$. The statement {\bf s1} is now
  proved upon replacing $p$ and $q$ by $p_i$ and $p_j$. The 
  statement {\bf s2} is true because of the structure of the matrix $diag
  [S_i]$ and the node indexing scheme described immediately before~\eqref{eq:lemma-statement}.

From Proposition~\ref{prop:union_graph_connected}, we have that
$\cup_{i=1}^{N}\vec{G}(\bar{F}_i)$ is connected. Since $S_i$ is the
off-diagonal part of $\bar{F}_i$, $\cup_{i=1}^{N}\vec{G}(S_i)$ is
connected as well.  Therefore, there is a path
from an arbitrary node $d_\kappa$ to another arbitrary node $d_\nu$ in
$\cup_{i=1}^{N}\vec{G}(S_i)$, for all $\kappa, \nu$ in $\{1,\dots, n^2\}$.
To prove the statement~\eqref{eq:lemma-statement}, pick such a path
from the node $d_\kappa$ to the node $d_\nu$ in $\cup_{i=1}^{N}\vec{G}(S_i)$, where
each edge in the path may lie in any of the graphs
$\{\vec{G}(S_i)\}_{i=1}^{N}$. For the sake of concreteness and
compactness, let us consider a path of length two, consisting of
the two edges $d_\kappa \rightarrow d_h$ and $d_h \rightarrow d_\nu$, which belong to the
graphs, say, $\vec{G}(S_\ell)$ and $\vec{G}(S_m)$, respectively. From 
{\bf s1} we have proved above, we know that there is a path from the node
$(p_i,d_\kappa)$ to the node $(p_\ell,d_\kappa)$ in the graph $\vec{\G}(\MarkovP^T
\otimes I)$, call this path $path[(p_i,d_\kappa) \leadsto
(p_\ell,d_\kappa)]$. From {\bf s2}, we have that the edge $(p_\ell,d_\kappa)\rightarrow (p_\ell,d_h)$ exists in the graph $\vec{\G}(diag[S_i])$ due to the existence of the edge $d_\kappa\rightarrow d_h$ in
$\vec{\G}(S_\ell)$. Thus, we have the path from $(p_i,d_\kappa)$ to $(p_\ell,d_h)$ in
the combined graph $\{\vec{G}(\MarkovP^T)\otimes \vec{G}(I) \}\bigcup
\vec{G}(diag[S_i])$ by
joining the path $path[(p_i,d_\kappa) \leadsto
(p_\ell,d_\kappa)]$ with the edge $(p_\ell,d_\kappa)\rightarrow (p_\ell,d_h)$. Using this idea
repeatedly, we construct a path from  $(p_i,d_\kappa)$ to $(p_j,d_\nu)$ in $\{\vec{G}(\MarkovP^T)\otimes \vec{G}(I) \}\bigcup  \vec{G}(diag[S_i])$
as follows:
\begin{align*}
  path[(p_i,d_\kappa) \leadsto (p_\ell,d_\kappa)], & \quad \text{ in }  \vec{G}(\MarkovP^T)\otimes \vec{G}(I)\\
 (p_\ell,d_\kappa)\rightarrow (p_\ell,d_h), & \quad \in \vec{G}(diag[S_i])\\
path [(p_\ell,d_h) \leadsto (p_m,d_h)], & \quad \text{ in } \vec{G}(\MarkovP^T)\otimes \vec{G}(I) \\
(p_m,d_h)\rightarrow (p_m,d_\nu), & \quad \in \vec{G}(diag[S_i])\\
path[(p_m,d_\nu) \leadsto (p_j,d_\nu)], & \quad \text{ in } \vec{G}(\MarkovP^T)\otimes \vec{G}(I),
\end{align*}
where each $path[\cdot]$ exists due to the property {\bf s1} established
above, and each edge exists due to the property {\bf s2} as well as with the
assumed existence of the edges $d_\kappa\rightarrow d_h$ and $d_h\rightarrow d_\nu$ in
the union graph. This argument can be extended to a path of any length
between $d_\kappa$ and $d_\nu$ in the union graph $\cup_{i=1}^{N}\vec{G}(S_i)$. Thus, there is a path from $(p_i,d_\kappa)$ to
$(p_j,d_\nu)$ in $\{\vec{G}(\MarkovP^T)\otimes \vec{G}(I) \}\bigcup
\vec{G}(diag[S_i])$, which proves sufficiency.

(Not connected $\Rightarrow$ reducible): A
simple counterexample proves necessity. Construct a trivial Markov chain with a single
state: $\graphset = \{\G_1\}$ (so that $\MarkovP = 1$) where $\G_1$
is an $n$-node graph without a single edge.  Then $\mathcal{D} = F_1
= J_1 \otimes J_1 = I$, which is reducible.  \qed
\end{pf*}
}

\end{document}